\PassOptionsToPackage{table,xcdraw}{xcolor}
\documentclass[acmsmall]{acmart}
\AtBeginDocument{%
  }

\setcopyright{acmlicensed}
\copyrightyear{2018}
\acmYear{2018}
\acmDOI{XXXXXXX.XXXXXXX}
\acmConference[CSCW '26]{CSCW '26: The ACM Conference on Computer-Supported Cooperative Work and Social Computing}{October 2026,
  2026}{Salt Lake City, Utah}
\acmISBN{978-1-4503-XXXX-X/2018/06}

\usepackage{graphicx}
\usepackage{subcaption}
\usepackage{array}
\usepackage[utf8]{inputenc}
\usepackage{fontenc}
\usepackage{multirow}
\usepackage{hyperref}

\usepackage{amssymb}
\usepackage{csquotes}

\begin{document}

\title{We Need Granular Sharing of De-Identified Data—But Will Patients Engage? Investigating Health System Leaders' and Patients' Perspectives on A Patient-Controlled Data-Sharing Platform}



\author{Xi Lu}
\affiliation{\institution{University at Buffalo, SUNY}\country{USA}}
\email{xlu30@buffalo.edu}
\orcid{0000-0002-7868-273X}

\author{Di Hu}
\affiliation{\institution{University of California, Irvine}\country{USA}}
\email{dih11@uci.edu}
\orcid{0000-0002-2842-1478}

\author{An T. Nguyen}
\affiliation{\institution{Cedars-Sinai Medical Center}\country{USA}}
\email{an.nguyen@cshs.org}
\orcid{0000-0003-2408-6307}

\author{Brad Morse}
\affiliation{\institution{University of Colorado Anschutz Medical Campus}\country{USA}}
\email{brad.morse@cuanschutz.edu}
\orcid{0000-0002-1080-3961}

\author{Lisa M. Schilling}
\affiliation{\institution{University of Colorado Anschutz Medical Campus}\country{USA}}
\email{lisa.schilling@cuanschutz.edu}
\orcid{0000-0002-6878-189X}

\author{Kai Zheng}
\affiliation{\institution{University of California, Irvine}\country{USA}}
\email{kai.zheng@uci.edu}
\orcid{0000-0003-4121-4948}

\author{Michelle S. Keller}
\affiliation{\institution{University of Southern California}\country{USA}}
\email{mkeller5@usc.edu}
\orcid{0000-0002-8157-7586}

\author{Lucila Ohno-Machado}
\affiliation{\institution{Yale}\country{USA}}
\email{lucila.ohno-machado@yale.edu}
\orcid{0000-0002-8005-7327}

\author{Yunan Chen}
\affiliation{\institution{University of California, Irvine}\country{USA}}
\email{yunanc@ics.uci.edu}
\orcid{0000-0003-4056-3820}

\renewcommand{\shortauthors}{Xi Lu et al.}
\renewcommand{\shorttitle}{Perspectives on a Patient-Controlled Sharing Platform for De-Identified Medical Data}

\begin{abstract}
  Patient-controlled data-sharing systems are increasingly promoted as a way to empower patients with greater autonomy over their health data. Yet it remains unclear how different stakeholders, especially patients and health system leaders, perceive the benefits and challenges of enabling granular control over the sharing of de-identified medical data for research. To address this gap, we developed a high-fidelity prototype of a patient-controlled, web-based consent platform and conducted a two-phase mixed-methods study: semi-structured interviews with 16 health system leaders and a survey with 523 patient participants. While both groups appreciated the potential of such a platform to enhance transparency and autonomy, their views diverged in meaningful ways. Leaders viewed transparency and granular control through the lens of informed consent and institutional ethics, whereas patients interpreted these factors as safeguards against potential risks and uncertainties. Our findings underscore critical tensions such as individual control and research integrity. We offer design implications for building trustworthy, context-aware systems that support flexible granularity, provide ongoing benefit‑centered transparency, and adapt to diverse literacy and privacy needs.
\end{abstract}

\setcopyright{cc}
\setcctype{by}
\acmJournal{PACMHCI}
\acmYear{2026} \acmVolume{10} \acmNumber{2} \acmArticle{CSCW043}
\acmMonth{4} \acmPrice{} \acmDOI{10.1145/3788079}

\ccsdesc[500]{Human-centered computing~Human computer interaction (HCI)}
\ccsdesc[100]{Human-centered computing~Empirical studies in HCI}

\keywords{Health System Leader, Patient, Stakeholder, Medical Health Data, Data Sharing, Medical Data, De-Identified Data}

\received{May 2025}
\received[revised]{November 2025}
\received[accepted]{December 2025}

\maketitle

\section{Introduction}
\label{sec:intro}

De-identified patients' electronic health records (EHRs), such as demographics, clinical notes, laboratory tests, and billing information, are often used in research to advance medical knowledge, improve treatment plans, and develop new medications and innovative technologies \cite{Morse2023,Kim2015, Benevento2023}. According to policies and laws, such as the Health Insurance Portability and Accountability Act (HIPAA) \cite{USDepartmentofHealthandHumanServices2025}, de-identified medical data can be shared and used for research without explicit patient consent as these data have been traditionally considered low risk \cite{Sandy2021, Dinh-Le2019}. 

However, current practices that permit the use of de-identified medical data without consent are increasingly being questioned \cite{murdoch2021privacy}. With advancements in computational techniques, especially artificial intelligence (AI) and large-scale data linkage, the risk of re-identification has increased \cite{packhauser2022deep, murdoch2021privacy}. While few studies directly examine patients' views on de-identified data use, existing research shows that many patients desire greater control over how their medical data are shared \cite{Morse2023, Kalkman2019}. Patients often report that current consent processes lack transparency and fail to provide options for selectively sharing specific data types (e.g., demographics, clinical conditions, genetic information) or limiting access to certain types of organizations \cite{Grande2014, Morse2023}. These concerns are compounded by patients' skepticism about how securely their data are handled, especially in an era where full anonymity is increasingly difficult to guarantee \cite{trinidad2020public, murdoch2021privacy, chiruvella2021ethical}. Collectively, these studies point to the strong preferences patients have in gaining more granular and transparent control over their de-identified medical record sharing \cite{Morse2023, Kim2015}, and suggest that patient-controlled sharing is necessary to balance the needs of enhancing medical research and respecting patients' preferences. 

While the past HCI and CSCW literature has extensively explored patient experiences and preferences in sharing their health data \cite{Weng2019, Kim2015, Morse2023, Chung2016, Pina2017,Murnane2018}, these studies have rarely focused explicitly on de-identified data and have primarily centered on patients' perspectives. Designing and implementing a system that allows patients to granularly control the sharing of their de-identified data requires buy-in from healthcare organizations. It is therefore necessary to consider the perspectives of health system leaders, whose experiences and insights into institutional culture and medical research are important for enabling patient-controlled data-sharing platforms. Previous HCI studies on examining the use of EHR highlight the importance of understanding the diverse perceptions of different stakeholders, as they often have distinct opinions and preferences \cite{Cajander2019, Bossen2012, Tang2015}. Understanding the perspectives of key stakeholders helps strike a balance between patients' needs and concerns with the operational and research needs of medical institutions. It can also provide valuable insights into how to integrate designs that empower patients' data-sharing control into existing health systems, as successful implementation requires addressing different stakeholders' practical considerations.  

In this study, we investigate health system leaders' and patients' perspectives of a patient-controlled system for sharing de-identified medical data for research, aiming to answer the following research questions:

\begin{itemize}
    \item RQ1: What are the current practices adopted by medical institutions for obtaining de-identified patient medical data and enabling researcher's access?   
    
    \item RQ2: What are health system leaders' perceived benefits and challenges of a patient-controlled system for sharing de-identified medical data for research?

    \item RQ3: What are patients' perceived benefits and challenges of a patient-controlled system for sharing de-identified medical data for research? 
\end{itemize}

To answer these questions, we developed a high-fidelity prototype to showcase a granular way to give patients control around their de-identified medical data, informed by existing research highlighting patients' concerns about the lack of transparency and granularity in current consent methods for medical data sharing \cite{Morse2023, Kim2015, Grande2014}. We then utilized this prototype as a design probe in a two-phase mixed-methods study, prompting health system leaders' and patients' perceptions. In the first phase, we conducted semi-structured interviews with 16 health system leaders, including chief medical/nursing information officers and CIOs, IRB and Research Compliance experts, and other research leaders. In the second phase, we surveyed 523 patient participants who interacted with the prototype before responding to measures on privacy concerns, health literacy, and willingness to use the system. This multi-stakeholder approach allowed us to compare how institutional decision-makers and patients perceive the benefits and risks of patient-controlled platforms for sharing de-identified health data. While leaders appreciated the increased autonomy and transparency the platform offers, they voiced concerns about its potential impacts on research quality and equitable participation. Patients generally supported the platform's empowering features but evaluated it through a lens of privacy, risk mitigation, and perceived personal benefit. This comparative perspective reveals critical tensions and design challenges in implementing such platforms at scale.

\vspace{0.3cm}
This study contributes to the field of HCI and CSCW in several ways: 
\vspace{-0.1cm}
\begin{itemize}
    \item An empirical understanding of the current practices adopted by medical institutions for obtaining de-identified patient data and enabling researcher access.     
    
    \item An empirical understanding and comparison of patient and health system leader perspectives on a system designed to provide patients with granular and transparent control of their de-identified medical data sharing.   

    \item Implications for carefully \textcolor{black}{designing patient-controlled data-sharing systems that support flexible, low-burden granularity, provide ongoing benefit-centered transparency, and offer literacy-adaptive consent experiences}.
  
\end{itemize}

\section{Related Work}
\label{sec:related-work}

\subsection{Studies on Sharing Medical Record Data for Research}

Medical records data has been widely used for health research, including demographics, clinical diagnoses, laboratory tests, and billing information \cite{Morse2023, Benevento2023}. The motivations for utilizing patients' medical data for research stem from both individual and societal benefits. For individuals, sharing personal health data for research can help increase one's medical knowledge \cite{Morse2023, Kalkman2019}. From a research perspective, access to patients' personal health data allows healthcare providers and researchers to improve diagnosis and treatment \cite{Kalkman2019}, provide personalized medicine \cite{Abul-Husn2019}, and create new knowledge \cite{Ross2014}. For instance, medical record data serves as a valuable source for developing AI models that can predict disease risks, outcomes, and treatment responses, advancing medical research and supporting clinical decision-making \cite{Benevento2023, Majnarić2021, Ross2014}. On a societal level, studies show the importance of collecting patients' health data during public health crises, such as the COVID-19 pandemic, to develop effective drug and vaccine treatments \cite{Benevento2023}, and make informed decisions about policies \cite{Dagliati2021}.

Recent advancements in Electronic Health Records (EHRs) and digital health systems have further increased the accessibility of medical record data for researchers. Researchers can access patients' medical record data in several ways, such as submitting research proposals to an Institutional Review Board (IRB), utilizing institutions' de-identified dataset platforms, and obtaining direct patient consent usually collected during clinical encounters \cite{Lanier2018}. 
Policies or laws require researchers and health institutions to obtain patient consent before using or disclosing their medical record data, typically through an opt-in or opt-out approach, aiming to respect patient autonomy while facilitating research participation \cite{Sandy2021, Morse2023}. In an opt-in approach, patients must actively signal their willingness to have their data included in research activities or shared with third parties \cite{Junghans2005}. In contrast, the opt-out approach assumes patient participation by default unless they explicitly refuse to participate in the research \cite{de_Man2023}. Studies show that most health organizations adopt the opt-out method to reduce potential selection bias among research participants while giving them choices and ensuring minimal burdens on both researchers and patients \cite{Sandy2021, Morse2023}. Additionally, in the U.S., the Federal Policy for Protection of Human Subjects allows waivers of informed consent for low-risk research \cite{Morse2023}. 

While many patients are willing to share their health data for research, a plethora of studies find their willingness is influenced by a range of factors, such as the intended data usage, the sensitivity of data types, the types of organizations requesting data, and patients' demographics \cite{Grande2014, Benevento2023, Weng2019, Kim2015}. For instance, patients are more willing to share their data for research projects or clinical trials, while their willingness decreases if the data is used for commercial purposes \cite{Grande2014, Benevento2023}. Patients perceive varying levels of sensitivity with different types of medical record data, with studies showing that individuals are generally more reluctant to share psychiatric information, substance abuse records, and domestic violence data compared to other types of data \cite{Weng2019, Weitzman2012, Benevento2023, Bourgeois2008}. Patients have varying trust towards different types of organizations, with studies showing that they are more likely to share their data with public health authorities and research institutions than commercial entities \cite{Morse2023, Weitzman2012, Kalkman2019}. Individuals' demographic factors (e.g., gender, age, education, and race) also play a significant role in determining data-sharing preferences \cite{Weng2019, Benevento2023, Sandy2021, Kalkman2019}. 

Prior studies further surface patients' concerns toward the current consent practices. Patients feel that they lack control over their data, including which studies it is shared with, who has access to it, and how it is used \cite{Morse2023, Kalkman2019}. More specifically, patients desire granular control over their data-sharing preferences, seeking to withhold certain sensitive data types (e.g., mental health data, sex orientations, and substance abuse) and choose which entities can access their data \cite{Morse2023, Caine2013}. Individuals have also raised concerns about the transparency in the data-sharing process \cite{Morse2023, Grande2014, Kalkman2019},  wanting to get more details about the broad consent that they are asked to sign \cite{Morse2023}. Privacy and data security are also important issues that influence individuals' willingness to share data \cite{Kim2015, Benevento2023}. Patients express concerns about the security of their EHR data and worry about potential misuse by third parties \cite{Morse2023}. 

\textcolor{black}{Despite extensive work on patient consent and data-sharing preferences, prior research has largely focused on identifiable health data, as regulations permit the use of de-identified data without explicit consent \cite{USDepartmentofHealthandHumanServices2025}. This paper addresses a key gap by examining patients’ nuanced preferences for sharing de-identified medical data. Specifically, we surface how factors like privacy concerns, perceived user burdens, and health literacy levels shape their interest in using patient-controlled platforms to manage and share such data for research.}

\subsection{HCI Research on Utilizing Health Data}

In recent years, the field of Human-Computer Interaction (HCI) has extensively explored the use of health data. While some research focuses on the use of medical data generated in clinical settings \cite{Cajander2019, Tang2015, Veinot2010}, the majority of research concentrates on personal health data collected through tracking and wearable devices in people's everyday lives, mainly around individuals' sharing health data with other stakeholders (e.g., healthcare professionals, caregivers, family members, and peers) \textcolor{black}{\cite{Lai2017,Chung2016,Pina2020,Chung2017,Gui2017}}.

HCI works on medical data have explored the use of medical record data, as well as the design and evaluation of relevant technology \cite{Sepehri2023, Pater2024, Iott2019, Lv2017, Cajander2019, Bossen2012, Marathe2021, Murphy2017, Tang2015, Veinot2010, Yoo2019, Dinh-Le2019, Mosaly2016}. For instance, although electronic health record (EHR) systems have traditionally been viewed as clinical tools for healthcare providers, recent works have explored the design of patient-accessible EHRs, allowing patients to view, manage, and contribute personal health data (e.g., diet, sleep, exercise, symptoms, and bio-metrics data) to their medical records \cite{Cajander2019, Sepehri2023, Tang2015, Veinot2010, Dinh-Le2019, Lv2017}. Making medical data accessible can help patients better understand their health conditions, improve patient-provider communications, and facilitate care coordination between multiple caregivers \cite{Ferreira2007, Sinha2021, Bourgeois2008}. Studies also point out the importance of investigating various stakeholders' perceptions of EHR, as stakeholders could have different opinions and preferences. For example, Cajander et al. \cite{Cajander2019} contrasted the perspectives of patients and healthcare providers on patients' access to EHR, with patients seeing the access as a way to check for errors and ensure accuracy, while providers suspecting whether patients derive meaningful information from their EHRs. Bossen et al. \cite{Bossen2012} critiqued that most studies focus on the collaborative use of EHR between patients, caregivers, and healthcare professionals, while giving limited attention to non-clinical staff. Tang et al. \cite{Tang2015} reported how the deployment of a EHR system disrupted the workflow of volunteers who supported the clinic, as the EHR was not designed for resourced-restricted volunteer-based organizations. 

Regarding personal health data, the HCI community primarily examines individuals' sharing personal health data (e.g., symptoms, food, diet, exercise, sleep, moods, and period data) with various stakeholders \cite{Epstein2020, Gui2017a, Chung2017, Figueiredo2020, Chung2016} and public health authorities \cite{Lu2021a, Arzt2021}. Patients frequently share health data with caregivers and family members \cite{Figueiredo2021,Murnane2018}, or allow them to directly track and collect data when patients are unable to do so on their own \cite{Sepehri2023, Jo2022}. This practice helps caregivers and family members gain an understanding of and foster empathy toward the patients' conditions \cite{Yamashita2017, Pina2017}. In addition to sharing offline with existing social networks, such as family members, friends, and providers, patients are increasingly sharing personal health data with like-minded audiences through online health communities and social media (e.g, Facebook groups, Reddit, and TikTok) \cite{Rubya2017, Huh2015, Massimi2014}. Besides getting emotional support \cite{Young2019, Mittal2023, Wang2023, Zhang2023}, studies find that peer-to-peer health information exchange empowers patients by fostering a sense of self-advocacy and agency, since patients can learn from each other and use the knowledge gained from peers to communicate with providers and push for additional diagnostics or treatment plans, especially when they feel marginalized by the traditional patriarchal healthcare systems \cite{Young2019, Massimi2014}. Existing studies also explore the collaborative management of personal health information (e.g., health history, symptoms, lifestyle routines, and bio-metric data) between patients and healthcare professionals \cite{Lai2017, Figueiredo2020}. These studies highlight several benefits of sharing personal health data with providers: empowering patients by giving them data control and becoming an active contributor to their health records, motivating patients to manage and record symptoms, and enhancing communications between patients and healthcare providers \cite{Chung2016, Figueiredo2020, Grimme2024}. As for providers, they could gain an in-depth and accurate understanding of patients' conditions \cite{Lai2017, Kim2024, Tadas2023, Hong2018}.

Besides individual-level health data sharing, the HCI community has also explored how personal health data can contribute to public health and societal good. Aggregated data from self-tracking apps, wearable devices, and mobile health platforms have been used to monitor community health trends and track disease spreads \cite{Lim2019, Freifeld2010}. During the COVID-19 pandemic, many studies explored factors influencing people's willingness to share their personal health data for the greater societal good \cite{Utz2021,Seberger2021,Diethei2021,Jamieson2021,Lu2021a}. 

\textcolor{black}{Current HCI research primarily investigates patients' data-sharing preferences and concerns in everyday contexts, while offering less attention to how systems could support patient control and agency over de-identified data in clinical research settings. This paper addresses this gap by investigating how patients perceive granular control over de-identified medical data for research, and extends HCI research by emphasizing ongoing, benefit-centered transparency as key to encouraging patient engagement in platforms that support data sharing for the public good.}

\subsection{Concerns and Tensions in Sharing Medical and Health Data}

Despite the benefits of sharing individuals' personal health and medical data, existing studies have identified several issues and concerns around their privacy and data-sharing autonomy. Studies found that individuals have nuanced privacy concerns when sharing personal health data with different stakeholders (e.g., partners, family members, friends, and healthcare providers), influenced by the data sensitivity and relationships between the sharer and the data receivers \cite{Lu2024b, Chung2016, Pina2020, Rubya2017}, especially when the data can reflect or indicate one's detailed lifestyle \cite{Epstein2013,Murnane2018}. For instance, pregnant women emphasize the importance of maintaining control over data sharing to protect their bodily autonomy and prefer to selectively share sensitive symptoms, such as bleeding, with partners and providers rather than parents, family members, and friends \cite{Lu2024b}. 
While individuals tend to trust their providers, they still express privacy concerns that they may lose data control and third parties may access to their data after the data is shared with the medical entity \cite{West2018, Oh2022, Chung2016}. 

This concern is echoed in the broader use of tracking and other health-related technologies. Studies have uncovered a wide range of privacy issues in such tools \cite{Alhajri2022, Robillard2019, Obar2020}, including the lack of readability in apps' privacy statements, making users hard to understand them \cite{Alhajri2022, Robillard2019, Obar2020}; the sharing of users' data with 3rd parties without individuals' consent \cite{Alhajri2022, Hutton2018, Robillard2019, Nguyen2022}; the reliance on one-time consent or even no consent or user agreements \cite{Alhajri2022, Hutton2018, Nguyen2022,Robillard2019}; the limited access for users to their own health data \cite{Hutton2018}; and poor compliance with legal regulations such as European Union's General Data Protection Regulation (GDPR) \cite{Nguyen2022, Mehrnezhad2020}. Even for apps that comply with privacy regulations like GDPR, studies have found that users often bypass privacy policies and terms of agreements due to the information overload these documents present \cite{Ajana2020, Obar2020}.

In medical setting, privacy concerns are a key barrier to patients' data sharing \cite{abdelhamid2017putting}. While studies indicate that patients may choose to share their medical data when they perceive personal or societal benefits \cite{perera2011views, mello2018clinical}, or when anonymity is believed to be preserved through de-identification \cite{whiddett2006patients, goodman2017identified}, many still express hesitation, particularly when sensitive information is shared beyond their healthcare systems, even for research purposes \cite{perera2011views}. Patients worry about data misuse and breaches that could lead to potential consequences, such as their data being exploited by commercial entities or used against them \cite{Kalkman2019, platt2015public, mello2018clinical}. 
For example, one study reported that patients expressed fears about the possibility of researchers being compelled to disclose their information to federal agencies\cite{goodman2018comparison}, while another indicated that patients raised concerns about whether adequate systems are in place to protect their data\cite{goodman2017identified}.

These fears are justified. Under HIPAA, de-identification involves removing 18 specific identifiers (e.g., names, addresses, medical record numbers, dates), thereby allowing for unrestricted use and sharing \cite{USDepartmentofHealthandHumanServices2025}. However, this approach does not guarantee anonymity, as studies repeatly reporting successful patient re-identifications using datasets deemed de-identified according to HIPAA or institutional standards \cite{el2011systematic}. The increasing use of advanced AI algorithms has heightened this risk. AI systems are capable of uncovering latent patterns in large datasets and linking data points from various sources, increasing the likelihood of re-identification\cite{murdoch2021privacy}. For example, one study demonstrated that a deep learning system could match X-rays taken over a decade apart and recover patient identities from de-identified X-ray images based on biometric features that persist over time \cite{packhauser2022deep}. 
Moreover, as healthcare systems increasingly partner with third-party companies to implement AI and other novel data analytic tools to improve patient care and clinical efficiency, patient data may be accessed and stored in external databases, exacerbating data security risks \cite{trinidad2020public, murdoch2021privacy, chiruvella2021ethical}. 

Finally, there is a broader tension between individuals' privacy concerns and collective benefits. Research on people's adoption of contact tracing technology during COVID-19 shows that the privacy and data security concerns influences people's willingness to share personal health data \cite{Lu2021a,Arzt2021}, and these concerns mainly stem from mistrust of the government \cite{Altmann2020,Zimmermann2021,Grekousis2021} and worries about data being shared with third parties like tech companies \textcolor{black}{\cite{Lu2021a,Mehrnezhad2022}}. Studies also suggest that sharing data with public health authorities can diminish individuals' sense of autonomy, as they may feel pressured by the expectations of government and the society \cite{Lu2021a, Lu2022}. However, some works reveal that individuals who prioritize the common good, such as containing the spread of the virus, tend to overlook privacy and data security concerns, willingly contributing their personal health data to support public health efforts \cite{Seberger2021, Diethei2021}.

In summary, existing studies on medical data sharing have largely focused on understanding patients' preferences, particularly regarding identifiable personal health data. There is limited insight into how de-identified data are perceived and managed by individuals. 
Additionally, the perspectives of health system leaders, who influence consent policies, manage data infrastructure, and oversee data access for research, remain insufficiently explored. This paper addresses these gaps by comparing the views of both patients and health system leaders on a granular and transparent consent approach, contributing new directions for designing patient-controlled systems that support the secondary use of de-identified medical data. 

\section{Methodology}
\label{sec:methods}

This study is part of a broader research project that proposes a low-cost system to support patient consent and decentralize data delivery and access across a network, while enhancing operational transparency. 

This study employed a mixed-methods approach, including semi-structured interviews with 16 health system leader \textcolor{black}{participants} and a survey study with 523 valid patient participants. The study was designed in two phases to build a comprehensive understanding of different stakeholders' perspectives on a patient-controlled data-sharing system for research using de-identified data. 

In the first phase, we conducted a semi-structured interview study with a high-fidelity prototype, as a design probe, to prompt health system leader \textcolor{black}{participants}' perspectives on patient-controlled systems for sharing medical data in the EHR system for research. The prototype consists of features designed to offer patients a granular and transparent way to manage their data-sharing preferences for medical research requests, inspired by existing studies that highlight patients' desire for greater granularity and transparency in the consent process \cite{Morse2023, Kim2015}. Given the limited nature of health system leaders, as well as their time-constrained roles, we chose semi-structured interviews to capture rich, in-depth insights that would be difficult to obtain through broader methods, such as surveys.

After gaining insights into leader \textcolor{black}{participants}' perspectives, we designed a large-scale survey study with patients to examine their attitudes and willingness to use such a system. Deploying the patient survey after the leader interviews allowed us to refine our inquiry to probe whether patients' perspectives aligned with, diverged from, or supplemented those of institutional leaders. In this second phase, patient participants were first invited to interact with the prototype to simulate a real-world experience and record their data-sharing preferences across different types of de-identified data. They then completed a survey that captured their views on the platform, along with measures of their privacy concerns, health literacy, and demographic background. The use of a survey enabled us to reach a large and diverse sample of patients, providing breadth and generalizability to complement the depth of the earlier interview findings from health system leader \textcolor{black}{participants}.

This study received approval from our university's Institutional Review Board (IRB). \textcolor{black}{We refer to our study participants as \textit{leader participants} and \textit{patient participants}. We use \textit{leaders}, \textit{patients}, or \textit{participants} more broadly when referring to stakeholder groups or both collectively, depending on the context.}

\subsection{High-Fidelity Prototype}

We designed a high-fidelity digital prototype of a patient-controlled platform for sharing de-identified medical data, using it as a probe to elicit health system leader \textcolor{black}{participants}' perspectives on granular and transparent consent, and to allow patients to simulate the experience of using such a system. The prototype contains four key features for patients: setting up granular data-sharing preferences, reviewing and deciding on study requests, sharing reasons for rejecting study requests or withholding specific data types, and viewing and modifying enrolled studies. These features are designed to envision scenarios that might empower patients with greater autonomy and control over their data sharing. We present the key features related to the granular consent process.

\subsubsection{Set Up A Granular Default Opt-in/Opt-out Preferences}

\begin{figure}[t]
    \footnotesize
      \centering
          \includegraphics[width=0.98\linewidth]{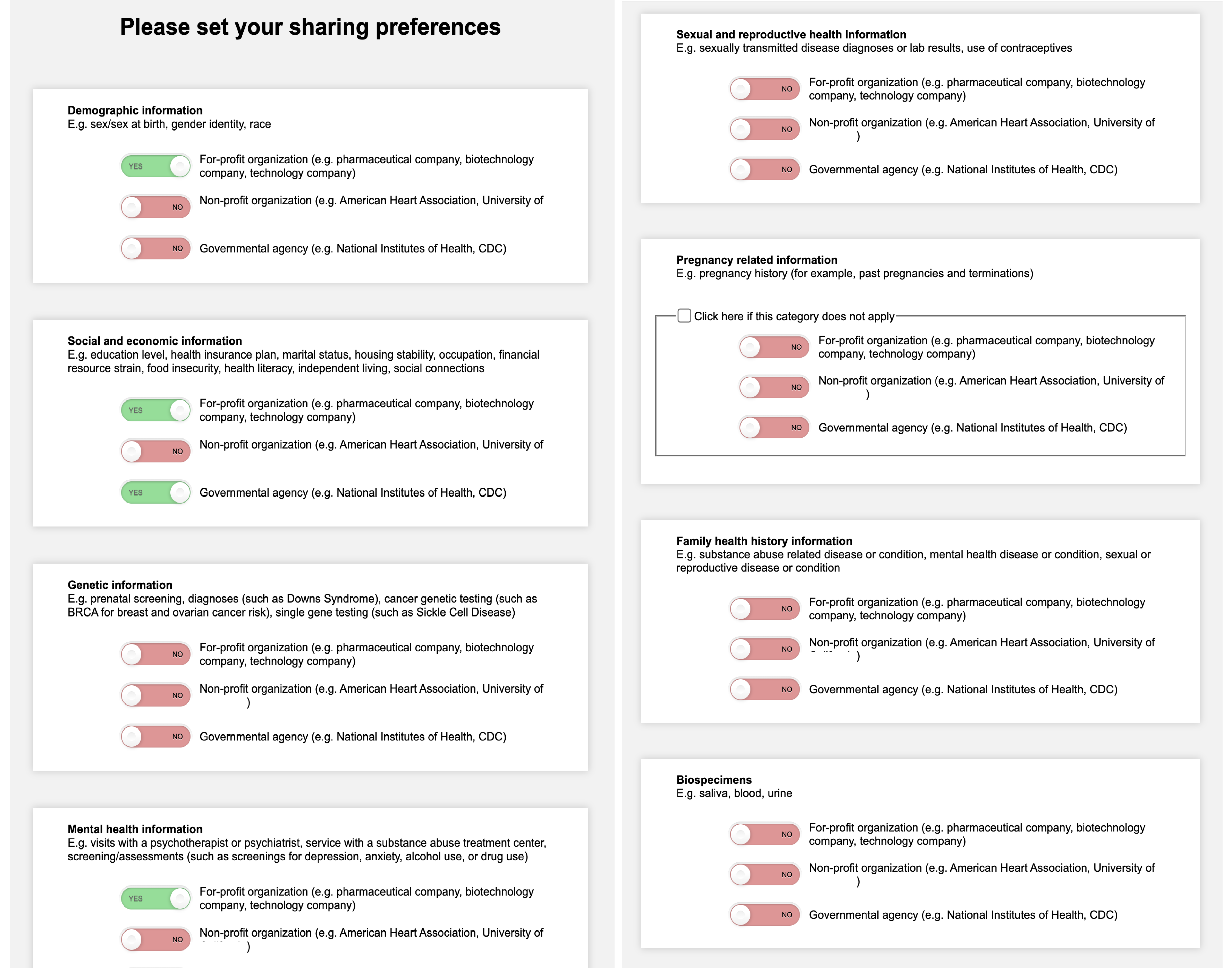}  
    \caption{The interface allows patients to have a granular way to set up default opt-in/opt-out for different data types (e.g., demographic information, social and economic information, genetic information, and mental health information). }
        \label{fig:default_sharing}
\end{figure}

As prior studies suggest that patients' willingness to share personal health data varies depending on the type of data \cite{Weng2019, Weitzman2012, Benevento2023},  we envision a scenario in which technology could support patients in making granular data-sharing decisions by allowing them to personalize sharing preferences for different data types. The prototype therefore enables patients to set up default opt-in/opt-out preferences among 10 different personal health data types, with detailed examples provided for some selective data types (Figure- \ref{fig:default_sharing}). We particularly include sensitive and controversial personal health data types, such as mental health information, pregnancy-related information, and sexual and reproductive health information, to probe how participants might perceive technology that could enable patients to withhold such sensitive health information.

The list of medical data types included in the prototype: 
\begin{itemize}
    \item Demographic information (e.g., sex/sex at birth, gender, and race) 
    \item Social and economic information (e.g., education level, health insurance plan, and marital status) 
    \item Genetic information (e.g., prenatal screening, diagnosis, and gene testing)
    \item Mental health information (e.g., visits with a psychotherapist or psychiatrist and screening/assessments)
    \item Sexual and reproductive health information (e.g., sexually transmitted disease diagnoses and lab results)
    \item Pregnancy-related information (e.g., past pregnancies and terminations)
    \item Family health history information (e.g., substance abuse-related disease or condition, mental health disease or condition, and sexual or reproductive disease or condition)
    \item Biospecimens (e.g., saliva, blood, and urine) 
    \item Immunizations (e.g., COVID-19, Influenza, and Measles) 
    \item Other general clinical information (e.g., diagnoses such as diabetes and hypertension, vital signs such as blood pressure and heart rate, medications, and lab results such as Xrays and CT scans) 
\end{itemize}

Since patients tend to have varying levels of trust in different organizations - non-profit organizations, for-profit organizations, and governmental agencies \cite{Morse2023, Weitzman2012, Kalkman2019}, the probe also enables patients to set up sharing preferences with different organizations for one specific data type (Figure 1):

\begin{itemize}
    \item For-profit organizations (e.g., biotechnology company, technology company, and pharmaceutical company)
    \item Non-profit organizations (e.g., American Heart Association and university)
    \item Governmental agency (e.g., National Institutes of Health and CDC)
\end{itemize}

\subsubsection{New Studies Requesting Your Data: A Transparent and Granular Opt-in/Opt-out Approach on Specific Studies}

\begin{figure}[t]
    \footnotesize
      \centering
          \includegraphics[width=0.98\linewidth]{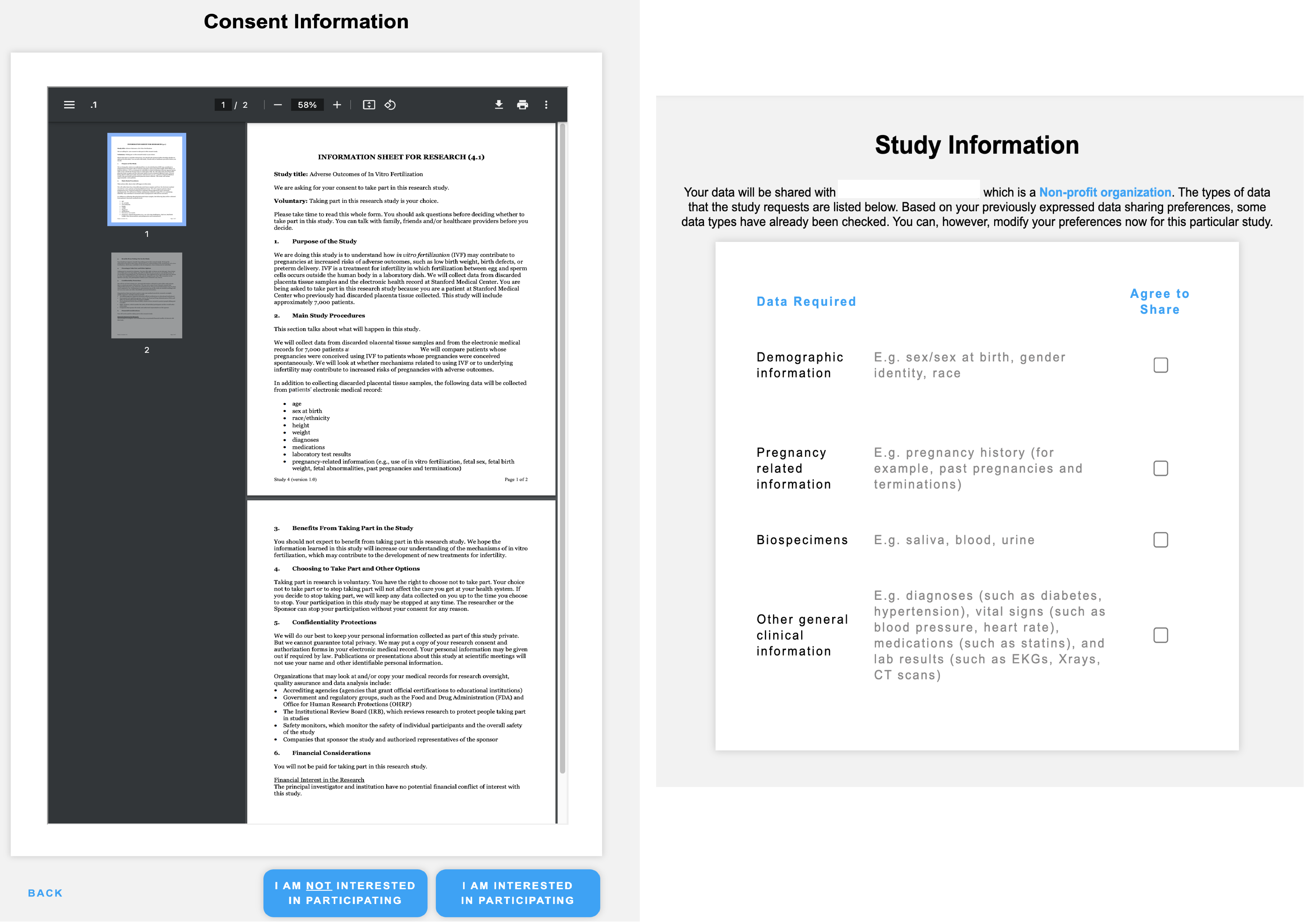}  
    \caption{If patients choose \textit{``I am interested''} for a specific study, they can view detailed consent information and set up data-sharing preferences for each data type that the study wishes to access. In the \textit{``study information''} page, the system also explains the requester's type (e.g., non-profit or for-profit organizations). }
        \label{fig:study_request}
\end{figure}

Existing studies indicate that patients wish for a transparent consent model, such as being able to learn more details about the broad consent that they sign in \cite{Morse2023}. Regarding this need, we propose that technology could empower patients by providing detailed information about any studies that request their data
. 
In the prototype, the \textit{``new data requests''} feature allows patients to: 1) provide a transparent way to review any studies requesting their data and access each study's consent information (Figure - \ref{fig:study_request} Left); 2) offer a granular option for patients to selectively share specific data types requested by a study, even after they have agreed to participate (Figure - \ref{fig:study_request} Right). 

\subsubsection{Elaboration: Sharing Reasons for Rejecting Requests or Withholding Data}

\begin{figure}[t]
    \footnotesize
      \centering
          \includegraphics[width=0.80\linewidth]{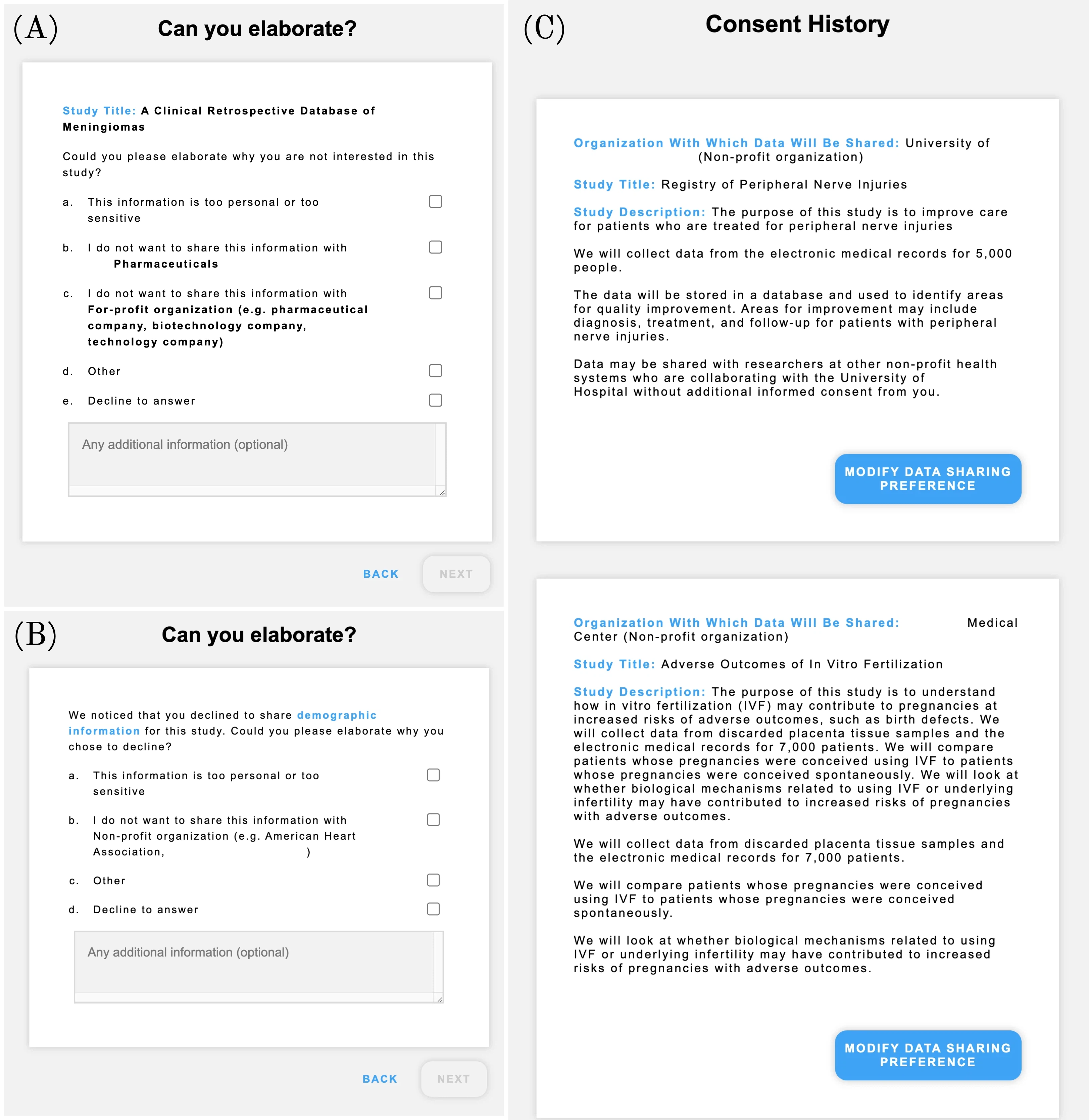}  
    \caption{\textcolor{black}{(A) If patients select “I am not interested” for a study, the system prompts them to specify their reason for opting out. (B) Patients can also state their reasons for withholding specific types of data when enrolling in a study. (C) The \textit{``Consent History''} interface. Patients can track all the studies that they have already opted in to and modify their existing data-sharing preferences for an enrolled study. Selecting \textit{``MODIFY DATA SHARING PREFERENCE''} leads to the \textit{``Study Information''} page (shown in Figure - \ref{fig:study_request}), where users can review and adjust the data previously shared.}}
        \label{fig:elaboration_view}
\end{figure}

Studies show that little is known about patients' reasons for opting out of a study \cite{Sandy2021}. We therefore propose that technology could allow patients to share their reasons when they opt-out of certain studies (Figure - \ref{fig:elaboration_view}A) or certain data types (Figure - \ref{fig:elaboration_view}B). \textcolor{black}{We also provide several options, with optional ``additional information'', to help patients explain their choices conveniently.} 

\subsubsection{\textcolor{black}{Consent History: Viewing and Modifying Enrolled Studies}}

\textcolor{black}{People’s data-sharing willingness can vary when situation changes \cite{Benevento2023}. Similarly, patients’ willingness to share personal health data for medical research may change over time, especially considering that medical research can sometimes last a long time. To accommodate this, the prototype includes a ``Consent History'' feature, allowing patients to monitor all their enrolled studies and modify data-sharing preferences for any specific study as needed (Figure - \ref{fig:elaboration_view}C). }

\subsection{Interview Process}

Our interviews sought to explore how health systems leaders–those who serve key roles in handling patients' medical data in a healthcare organization–perceive patient-controlled data-sharing platforms. The interviews were conducted between November 2023 and April 2024.

The interviews contained three parts. First, we asked participants about their institutions' consent processes and their roles relevant to this process, asking questions such as \textit{``What is your role in managing research-related data in your institution?''}, \textit{``What are your institution's current practices for using and sharing data collected from electronic health records and other medical record data for research studies?''}, and \textit{``Do patients opt in or opt out of sharing their data during their first healthcare encounter?''} Next, we presented the prototype as a probe to prompt participants' feedback about patient-controlled platforms designed to empower patient' autonomy through granular and transparent data-sharing features. We provided a demo of the prototype using a pre-recorded video, and also shared the key features in a Google Doc for participants to review at their convenience. Finally, after watching the video, we inquired participants' perceptions toward the system with questions like \textit{``What do you think about the concept of the prototype in general?''}, \textit{``What are the positive and negative aspects of the prototype?''}, and \textit{``In your opinion, will the use of the prototype change which patients are represented (or not) in research studies?''} All interviews were semi-structured, with each lasting approximately one hour.  

\subsection{Interview Participants}

We used a purposive sampling approach to get perceptions from people representing health system leadership who plays key roles in handling medical data, such as Chief Medical/Nursing Information officers (CMIO and CNIO) and Chief Information Officers (CIO) (implementer decision makers), IRB and Research Compliance experts (institutional representatives), and other research leaders (e.g., research infrastructure decision makers). 

We recruited participants through a combination of methods: using email lists from our network of research investigators, contacting potential participants listed on institutions' public websites, and employing snowball sampling after we interviewed some participants. We compensated each participant with \$75. 

\begin{table}[t]
	\centering
		\caption{Participants' Roles in Health Systems. }
	\label{tab:demo_leaders}
\resizebox{\textwidth}{!}{%
\begin{tabular}{rll}
\hline
\multicolumn{1}{c}{\textbf{ID}} &
  \textbf{Instituions and Type} &
  \textbf{Role} \\ \hline
1 &
  A, University hospital &
  CIO \\ \hline
\cellcolor[HTML]{FFFFFF}{\color[HTML]{222222} 2} &
  \cellcolor[HTML]{FFFFFF}A, University hospital &
  CMIO \\ \hline
3 &
  B, University hospital &
  CIO \\ \hline
4 &
  A, University hospital &
  Professor; Medical Oncologist \\ \hline
5 &
  \begin{tabular}[c]{@{}l@{}}C, Non-profit medical\\  teaching center\end{tabular} &
  Vice Dean for Research \& Graduate Education \\ \hline
6 &
  B, University hospital &
  Director of Enterprise Data \& Analytics \\ \hline
7 &
  \begin{tabular}[c]{@{}l@{}}C, Non-profit medical\\  teaching center\end{tabular} &
  \begin{tabular}[c]{@{}l@{}}IRB Director; Executive Director of the \\ Office of Research Compliance and Quality Improvement\end{tabular} \\ \hline
8 &
  \begin{tabular}[c]{@{}l@{}}C, Non-profit medical\\  teaching center\end{tabular} &
  Lead Data Intelligence Analyst \\ \hline
9 &
  \begin{tabular}[c]{@{}l@{}}C, Non-profit medical\\  teaching center\end{tabular} &
  Associate Dean for Research and Clinical Trials \\ \hline
10 &
  B, University hospital &
  Compliance office \\ \hline
11 &
  D, University hospital &
  IRB Assistant Director \\ \hline
\cellcolor[HTML]{FFFFFF}{\color[HTML]{222222} 12} &
  \cellcolor[HTML]{FFFFFF}D, University hospital &
  \begin{tabular}[c]{@{}l@{}}Professor; Associate Dean for Clinical Research, \\ Chief Clinical Research Officer\end{tabular} \\ \hline
\cellcolor[HTML]{FFFFFF}{\color[HTML]{222222} 13} &
  \cellcolor[HTML]{FFFFFF}D, University hospital &
  Senior Director of Research Administration \\ \hline
\rowcolor[HTML]{FFFFFF} 
{\color[HTML]{222222} 14} &
  {\color[HTML]{222222} E, University hospital} &
  {\color[HTML]{4D5156} Director of Research Informatics} \\ \hline
\cellcolor[HTML]{FFFFFF}{\color[HTML]{222222} 15} &
  \cellcolor[HTML]{FFFFFF}{\color[HTML]{222222} E, University hospital} &
  \begin{tabular}[c]{@{}l@{}}Associate Director of Compliance and Process Management \\ for the Enterprise Information \& Analytics department\end{tabular} \\ \hline
\cellcolor[HTML]{FFFFFF}{\color[HTML]{222222} 16} &
  \cellcolor[HTML]{FFFFFF}{\color[HTML]{222222} E, University hospital} &
  Sr. Data Compliance Specialist \\ \hline
\end{tabular}%
}
\end{table}

Table- \ref{tab:demo_leaders} describes the role of each participant at their institutions and their institutions' types. We did not collect participants' self-reported demographic information, such as gender, race, educational background, and income, as these factors were deemed to have minimal impacts on participants' perceptions of patient-controlled data-sharing platforms from the viewpoint of health system leaders. 

\subsection{Survey Design and Process}

The survey were first pilot tested with 19 participants. The final survey was deployed in June 2024. It was designed to mirror the themes explored in the leader interviews, focusing on capturing patients' attitudes and willingness to use the platform, their privacy concerns, and their level of health literacy. To achieve this, we combined validated instruments such as the HINTS Privacy Concerns scale \cite{nelson2004health, hesse2005trust} with self-designed close-ended and open-ended questions. Details on the survey design are provided in the following section.

After providing e-consent, participants accessed the prototype website via an internet browser. They were asked to interact with the prototype by setting their data-sharing preferences (yes / no) (Figure- \ref{fig:default_sharing}), and exploring preset study data request pages (Figure - \ref{fig:study_request}). Their data-sharing preferences during this process were recorded as system log data for subsequent analysis.

Following their interaction with the prototype, participants then completed a survey designed to capture their attitudes toward the platform, data privacy concerns, health literacy level, and demographic background.

The survey included the following items: 1) \textbf{Attitudes and willingness to use the platform}:  This section featured 7-point Likert scale items measuring participants' perceived usefulness and likelihood of using the platform (Table \ref{tab:atti_willing_platform}), as well as one open-ended question asking them to explain their willingness or reluctance to use it personally. Example questions included: \textit{`` If the platform was available to me where I receive health care, it would make me feel more confident that I have some say in who is allowed to collect, use, and share my medical information for research''}, \textit{``If the platform was available to me where I receive health care, I would personally use it ''} \textcolor{black}{followed by \textit{``Please briefly explain your response''}, and \textit{``How likely is it that you would recommend
iAgree to a friend or colleague?''} followed by \textit{``What is the main reason for your score?''}}; 2) \textbf{Attitudes and behaviors towards data privacy}, measured using the HINTS Privacy Concerns sub-scale score \cite{nelson2004health, hesse2005trust}. Example questions included: \textit{``How confident are you that you have some say in who is allowed to collect, use, and share your medical information?''}, \textit{`` How confident are you that safeguards(including the use of technology) are in place to protect your medical records from being seen by people who aren't permitted to see them?''}; 3) \textbf{Patients' health literacy level}, measured using the TOFHLA Health Literacy instrument \cite{parker1995test}. Example questions included: \textit{``How confident are you filling out medicalforms, such as forms that your doctor ask you to sign?''} and \textit{`` How often do you have someone help you read medical forms, such as hospital materials?''}; 4) \textbf{Demographic information}, including education level, employment status, annual household income, race and ethnicity, gender identity, sexual orientation, and age.

\subsection{Survey Participants}

We utilized multiple recruitment strategies to enroll study participants. These included outreach emails to individuals who had previously agreed to be contacted through the University C's Medical Center's Patient and Family Advisory Council; snowball sampling; social media posts on LinkedIn, X (formerly Twitter), and Facebook; and flyer distribution at collaborating institutions, including Universities A, B, C, D, and E. Participants received a \$10 gift card as compensation.

Prospective participants first completed a brief screening survey to assess eligibility, which included: (1) 18 years of age or older, (2) currently receiving healthcare in the United States, and (3) self-attestation that the participant could read English proficiently. Eligible participants were then provided a link to access the prototype platform, followed by the post-use survey. 

Table \ref{tab:demo_survey_patient} provides an overview of the final patient participants. After excluding 66 participants who failed attention checks, contained missing values, or had nonsensical answers to the open-ended questions, our final sample included 523 participants. Participants had a median age of 35 years (inter-quartile range:31.00–39.00). 69.21\% of participants identified White, 20.65\% identified as Black or African-American, 4.21\% identified as Asian, and 5.93\% identified as Other. Most were non-Hispanic (96.37\%), and held a bachelor's degree or above (44.55\%). The majority reported an annual income above \$50,000 (66.92\%).

\begin{table}
    \centering
    \caption{Demographics of Valid Survey Participants}
    \label{tab:demo_survey_patient}
    \begin{tabular}{|>{\raggedright\arraybackslash}p{0.5\linewidth}|c|} \hline 
    \multicolumn{1}{|c|}{\textbf{Demographics}} & \textbf{Overall (n = 523)}\\ \hline 
 Age, median (IQR)&35.0 [32.0, 39.0]\\ \hline 
         \multicolumn{2}{|l|}{Sex at Birth, n (\%)}\\ \hline 
         Female& 238 (45.51)\\ \hline 
         Male& 285 (54.49)\\ \hline 
 \multicolumn{2}{|l|}{Race and Ethnicity, n (\%)}\\ \hline 
 Asian&22 (4.21)\\ \hline 
 Black or African-American&108 (20.65)\\ \hline 
 White&362 (69.21)\\ \hline 
 Other&31 (5.93)\\ \hline 
 \multicolumn{2}{|l|}{Ethnicity, n (\%)}\\ \hline 
 Hispanic&19 (3.63)\\ \hline 
 Non-Hispanic&504 (96.37)\\ \hline 
 \multicolumn{2}{|l|}{Education, n (\%)}\\ \hline 
 College Degree or Higher&233 (44.55)\\ \hline 
 No College Degree&290 (55.45)\\ \hline 
 \multicolumn{2}{|l|}{Income, n (\%)}\\ \hline 
 \$50,000 or less&173 (33.08)\\ \hline 
 \$50,000 or more&350 (66.92)\\ \hline 
    \end{tabular}
    
\end{table}

\begin{table}
    \centering
    \caption{Attitudes and Willingness to Use the Platform}
    \label{tab:atti_willing_platform}
    \begin{tabular}{|>{\centering\arraybackslash}p{0.6\linewidth}|c|c|c|} \hline 
 Survey Questions& \multicolumn{3}{|c|}{Response, n (\%)}\\ \hline 
         &  Agree &  Neutral& Disagree\\ \hline 
         1. If the platform was available to me where I receive health care, I would personally use it.&  445 (85.09)&  32 (6.12)& 46 (8.79)\\ \hline 
         2. If the platform was available to me where I receive health care, it would make me feel more confident that safeguards are in place to protect my medical records from being seen by people who are not permitted to see them&  432 (82.60)&  36 (6.88)& 55 (10.52)\\ \hline 
 3. If the platform was available to me where I receive health care, it would make me feel more confident that I have some say in who is allowed to collect, use, and share my medical information for research.& 440 (84.13)& 28 (5.35)&55 (10.52)\\ \hline
    \end{tabular}
    
\end{table}

\subsection{Data Analysis}

\subsubsection{Qualitative Analysis}
All the interviews were conducted remotely via Zoom and recorded, with each session taking approximately one hour. We used Zoom's automatic transcription service with some necessary manual editing to ensure accuracy. We qualitatively analyzed the interview data by using the thematic analysis \cite{Braun2012}. The first and second authors first read five transcripts and open-coded them to generate some takeaways. They regularly held meeting to compare each other's insights, identifying some preliminary themes. We organized insights into two main categories: (1) current practices for obtaining patient consent adopted by participants' institutions and (2) participants' perspectives on the prototype. The first author then built a codebook with these two categories as the parent codes and utlized the codebook to code all the interviews. The codebook contained seven child codes: opt-in practice, opt-out practice, positive-granularity, positive-transparency, negative or tension-individual versus research, negative or tension-information burdens for general populations, negative or tension-negative impacts on certain populations. We applied a similar thematic approach to analyze responses of the open-ended questions in the patient survey.   

\subsubsection{Quantitative Analysis}
We applied descriptive statistics to report the sociodemographics of patient survey participants and their responses. Continuous variables were presented as median and interquartile range (IQR), while categorical variables were summarized using frequencies and percentages. Ordinal logistic regressions were employed to examine the relationship between patient characteristics and their willingness to use and attitudes (Table \ref{tab:atti_willing_platform}) toward our prototype. All responses were grouped into three levels of agreement: agree, neutral, disagree.
Additionally, we conducted Cochran's Q Test on the system log data, which captured patient participants' data-sharing preferences during prototype interactions, to assess whether their willingness to share differed significantly across various types of medical data. All quantitative analyzes were performed with R (Version 4.4.3).

We refer to leader participants as LXX and patient participants as PXX. 

\subsection{Limitations}

All participants were from large urban medical centers, which may limit findings' generalizability to health systems in rural setting. For example, rural health systems may have challenges such as limited access to new technology. Also, these systems may have a higher portion of patients with lower socio-economic status than non-rural institutions. The technological challenges that rural health systems face, combined with their patients' limited access to and interaction with technology, may lead to leader \textcolor{black}{participants} in these settings having different, potentially more critical views on our concept. In addition, since all participating institutions are academic medical centers, the results may not be representative of non-academic settings where research is not part of the organization's activities. Different priorities and operational contexts in such settings could influence leader \textcolor{black}{participants}' views on the need of requesting and getting patients' medical data for research. \textcolor{black}{In addition, while this study focused on patients and health system leaders, two stakeholder groups central to the deployment and adoption of patient-controlled data-sharing platforms, other key stakeholders like clinicians and researchers were not directly represented. Future studies could investigate the perspectives of clinicians and researchers, as clinicians may experience additional burdens if tasked with introducing or explaining such systems during time-constrained medical encounters, and researchers rely on patient data for their work.}

In this study, we utilized a high-fidelity prototype presented through a pre-recorded video for our demonstration to elicit participants' perceptions. While this approach enabled us to effectively present our concept, especially during remote interviews, it may have limited participants' understanding, as their perspectives was based on watching a demonstration of the prototype rather than interacting with it or using it in a real-world setting. Future research on the deployment of patient-controlled data-sharing platforms could provide valuable insights into both the benefits and challenges of implementing such systems.

\textcolor{black}{
For patient participants, the validity of our survey results may be limited by self-reporting and acquiescence bias. We acknowledge that this concern is especially relevant in the privacy-sensitive context of consenting to share electronic health records. Specifically, the statistical analysis based on participants' interactions with the prototype, including their chosen privacy settings, is subject to validity issues. Participants' responses to hypothetical questions about their future behaviors (e.g., whether they would use such a system if it was available to them) and their decisions within an experimental environment may not fully reflect their true preferences and actions in real-world settings. We therefore interpret these results as indicative of patients’ preliminary concerns and preferences regarding patient-controlled, granular de-identified data-sharing systems, rather than as precise forecasts of actual uptake or behavior. Future studies should extend this work by examining actual use, perceptions, and preferences across diverse health systems and stakeholder groups, including pilot implementations of granular data-sharing platforms in production environments.}
\section{Results}

\subsection{Current Practices for Obtaining De-Identified Patient Data and Enabling Researcher Access}


\subsubsection{General Opt-In at Initial Encounter While No Consent for Subsequent Use of De-Identified Data}

All the institutions that we interviewed were involved in data-oriented research, and generally asked for patients' consent for using their data in future research when first providing care to patients. For example, one institution had patients sign consent forms during the initial care encounter: \textit{``So there is a terms and conditions agreement that the patient sign, and it informs the patient that their data may be used for research. There's on that terms and conditions form. (L15)''}  L3's institution adopted a similar approach that \textit{``when you enroll in to receive any care, you sign some sort of paperwork. (L3)''} 

When it comes to the subsequent use of patients' de-identified data, leader participants noted that their institutions typically did not seek patient approval, as de-identified data was widely perceived as safe and posing minimal or no risk. L6 explained: \begin{displayquote}
\textcolor{black}{\textit{``if my data was de-identified and included in a data set that a researcher was looking at something that happened in the past again. Let's just take some sort of study for Covid, and I'm not identifiable. You know, it's gonna be used in aggregate. You're not gonna know who it is and so forth. The risk to me you know my feeling about it just personally is is very minimal, right? (L6)''}}\end{displayquote} Leader participants also emphasized the impracticality of re-contacting individuals for consent once data had been de-identified: \textit{``So for de-identified data, because we don't know who they are. So we don't typically get the consent. We get the HIPAA. Also additional labor. That's basically the say that tell IRB there's no way we can get the patient consent in. (L3)''}

\subsubsection{Self-Service Access to De-Identified Data: No IRB for Internal Use}

Institutions commonly de-identify portions of patients' medical data and store them in secure data warehouses, which researchers can access for analysis. As L10 introduced, \textit{``They [researchers] can use a data warehouse, which is a secure platform. They basically log in and they cannot download the data. They can go into a environment where they can correlate the data, and then they can get the data within the environment. They can use SAS, or python, whatever to analyze the data.''} 

Most institutions in our study did not require IRB approval for researchers using these de-identified datasets internally, as such use is typically not classified as human subjects research. L2 noted: \begin{displayquote} \textcolor{black}{\textit{``Self-service data does not require IRB, because it's been considered [as] not human subjects research. Those datasets are de-identified to limited dataset[s]. From a Hipaa perspective, they're provided in an environment that is closed. The users cannot move the data in, or they can move data in or data out. ''} This self-service model allowed internal researchers to quickly access data as "\textit{They don't need to do the IRB, and they just get an account. (L2)}"} \end{displayquote} However, when de-identified data was requested by external researchers, such as those from non-profit or for-profit organizations, additional review and formal agreements were typically required: \textit{``External entities, you have to do a data use agreement, and they would have to have an irb internally because they are not employees of the institution. (L4)''}   

\subsubsection{Limited Opt-Out Options in Some Institutions}

Compared to the general opt-in approach during the first care encounter, patients' ability to opt out of data usage was often limited, as only some institutions offered this option while others did not provide this choice at all. One leader participant said: \textit{``We don't have a [opt-out] procedure. You sign some [consent and agreement] paperwork. (L3)''} In comparison, some institutions enabled patients to opt out either verbally, filing out opt-out forms, or updating preferences on their EHR portals. For example, patients at one interviewed institution can \textit{``verbally opt out of the study itself. And then they have to do a written revocation of their authorization (L11).''} Another institution that used Epic—software that manages the medical records of 78\% of the U.S. patients \cite{Epic2024}—allowed patients to exclude their data from any research through a break-the-glass feature: \begin{displayquote}\textcolor{black}{\textit{``Break-the-glass [is] basically patients who have either they need their identity protected, or they opt into some special protections for the hospital. I don't know how the channel works for that, but there is an additional flag that we see on our end that says, Oh, this patient has been marked for special treatment and should not be included in any research. (L8)''}}\end{displayquote} Interestingly, in institutions providing the opt-out option, leader participants noted that only a few patients actually chose to exclude their data from research. L1 observed that most patients were inclined to contribute their data: \textit{``It's a small percentage. It'd be in the single digits from a percentage standpoint. I think most people are pretty liberal in that regard. (L1)''} At L8's institution, only \textit{``a quarter of a percent of patients''} opted out.

\subsubsection{Current General Consent Practices: All-or-Nothing and Lack of Transparency}

The existing consent format at all the institutions we interviewed followed an all-or-nothing approach, where patients must either opt in or opt out of their entire medical records for research. For instance, L1's institution was \textit{``kind of all or nothing (L1)''} that patients could not select which studies they wanted to participate in or exclude themselves from. Similarly, regarding the data types, patients had no choice in deciding which elements of their medical records they were willing to share. For example, one institution utilized a blanket consent method, which asked patients to consent to the use of their medical records data for any potential future studies without restrictions \cite{Caulfield2007}. L6 at this institution felt the blanket consent simplified the administrative burden that institutions faced when auditing and tracking research data involving patients' information: \textit{``The consent currently, by just having a blanket consent, It makes [things] easy. If you have allowed people to opt out. And I'm not saying that we shouldn't do this. I'm just talking about the challenges that this creates. You have to have a way of tracking that. Obviously, you have to have a way when you produce a data set of making sure that you exclude those people. and then you have to have a way of keeping that updated (L6)''}

Participants' institutions generally did not publicize to patients that their medical record data would be used for research: \textit{''here isn't a whole lot of institutions that will put a big banner up and say, Hey, you know, we're doing research with your data (L12).''} At institutions that allowed for opt-out, this option was often hidden from patients. For example, L15's institution did not explicitly inform patients of the opt-out option during their first encounter: \textit{`` There's that [consent] terms and conditions form. There is not an explicit opt-out option at this time. We do have a process where we can update the EMR [electronic medical record] for a patient if they decide that they don't want their data to be used for research but they're not explicitly asked during their first encounter in the patient care (L15)''} 

\subsection{Health System Leaders' and Patients' Perceived Benefits of Designing Patient-Controlled Systems for De-Identified Data sharing}

Following the previous section, which examined current institutional practices for obtaining de-identified patient data and enabling researcher access, this and the next section present the perceived benefits and challenges of a patient-controlled data-sharing platform for de-identified data. We first describe the perspectives of health system leaders, which offer detailed insights based on their institutional roles and decision-making responsibilities. We then present patient participants' perspectives, emphasizing how their views align with, add nuance to, or diverge from those of institutional leaders. 


\subsubsection{Health System Leaders' Perceived Benefits}

Overall, compared with traditional consent approaches, all the leader participants stated that they appreciated the greater levels of patient autonomy offered by a patient-controlled platform, as it could provide increased transparency and granularity in the data sharing process: \textit{``[patient-controlled platforms] definitely empower individuals to have full control over their data (L13). ''}  

\textbf{Granular Control Empowers Selective Data Sharing. }Leader participants liked our concept for its granular data-sharing control, feeling it could empower patients by being able to choose specific data requesters (e.g., non-profit institutions and for-profit institutions) and selectively share data. The health system leaders felt some patients might choose to opt out of sharing particular types of data due to specific sensitive information in their medical records that they preferred to keep private, while the traditional approach required them to either share all the data or opt out completely. For instance, studies show that general patients have greater privacy concerns with mental health information than non-psychiatric data, mainly due to its sensitivity and the associated stigmas \cite{Benevento2023,Clemens2012}. Therefore, a consent method that allowed for selective data sharing would be more beneficial. L1 said: \begin{displayquote}\textcolor{black}{\textit{``For the people who opt out there may be a reason why they're opting out. And it's because I have something that's really sensitive in a certain pocket of my medical record that I don't want people to see. and if I can exclude that section, whether it be mental health, or you know, a disease, you know, type or something. Then maybe I'm okay sharing my [other] general primary care data, right?  So I think it would be beneficial. (L1)''}}\end{displayquote} 
Leader participants also liked that the proposed system allowed patients to choose with which types of organizations to share their data, recognizing that patients may have different levels of trust and acceptance towards non-profit organizations, for-profit companies, and governmental agencies. L9 explained: \textit{``I think it's important cause it's not just do you wanna share your data, yes or no. That's what it's like at most places. This is much more granular because I may do a trial with the NIH, but I don't want to do a trial with industry. or I may want all of my records, but not my mental health records released. That's what I like about [it]. (L9)''} 

\textbf{Transparency Helps Patients Make Informed Decisions About Data Use. }Contrary to the existing consent approach, which often failed to explicitly inform patients about how their data will be used and sometimes obscured the opt-out option, leader participants perceived the proposed design as offering greater transparency, such as allowing patients to see which studies and organizations are requesting their data and review all the enrolled studies. With a platform where patients could control their enrollment for different studies, providing necessary information about the studies requesting their data (e.g., consent information and the types of organizations involved) could help patients make informed decisions. L10 emphasized that allowing patients to see where their data will be going could be a significant advantage of patient-controlled platforms, as people would be curious about how their data will be utilized: \textit{``Patients should be able to see where data is going, which study their data will be used. I think people are curious about that. (L10) ''} L3 said: \textit{``It's good for them [patients] to know whether it's a for-profit or non-profit [organization], because the [existing] study consent forms don't necessarily tell like [this]. They may say this, but it may not be very obvious. (L3) ''} 

\textbf{Capturing Patient Feedback Can Improve Study Design and Participation.} Leader participants also liked the proposed platform's feature which allowed patients to disclose their reasons for rejecting a study request or withholding specific data types for an enrolled study, as this could provide valuable insights for researchers to redesign studies or address patients' concerns in future research. L9 said: \begin{displayquote} \textcolor{black}{\textit{``There may be a specific reason [why patients do not participate], and [the design] allows you to have specific reasons why you're not in a study, but otherwise you could be in a study. So it allows us to have potentially greater enrollment in studies. (L9)''}}\end{displayquote} This suggests that by informing researchers of patients' concerns, future studies could be designed to address these issues, potentially leading to greater participation as patients might feel more comfortable with studies that take their concerns into consideration. 

\subsubsection{Patients' Perceived Benefits}

Overall, most patient participants expressed positive attitudes toward the proposed platform after interacting with it. Among 523 patient participants, 445 (85.09\%) agreed that they would personally use the platform if it was available to them when receiving health care (Table \ref{tab:atti_willing_platform}). A majority of participants agreed that the platform would enhance their sense of privacy and autonomy. Specifically, 432 (82.60\%) reported feeling more confident that the system would help safeguard their medical records from unauthorized access, and 440 (84.13\%) felt more confident in having control over who is allowed to collect, use, and share their medical information for research (Table \ref{tab:atti_willing_platform}). One patient explained: \textit{`` I would definitely use [the platform] as it aligns with my values of privacy and data protection, giving me a voice in the research process. (P241)''} 
The autonomy afforded by the platform led patient participants to see themselves as active collaborators in research rather than passive data sources, an experience often missing from traditional consent models. P331 appreciated the platform would \textit{``offer me the chance to collaborate with researchers and healthcare professionals for the greater good''}. 

\textbf{Patients Used Granularity to Manage Sensitivity and Address Uncertainty. }With regard to granularity, our Cochran's Q Test revealed significant variation in patient willingness to share different types of de-identified health data, regardless of the recipients (Q statistics: 117.70 for non-profit, 142.95 for for-profit, 186.20 for government agency; all p-values < 0.01). Higher Q statistics indicate more pronounced differences when sharing with government agencies. Log data showed that patients were least willing to share pregnancy-related information (below 40 \% for all organization types) and were particularly hesitant to disclose additional general clinical information to government agencies (36.14\%), while all other data types had over 50\% of participants willing to share. This aligns with leaders' observations that patients may want to withhold particularly sensitive information while still contributing less sensitive data. 

In contrast to leader participants, who generally viewed de-identified data as low-risk, patients showed more hesitation and uncertainty about whether de-identified data could truly protect their identities. For example, P26 asked: \textit{``So that means any of the demographic information that's shared has no link to my own identity. Right?''} Similarly, P27 expressed: \textit{``They [researchers] cannot find my actual like personal information''} if sharing genetic data. This suggests that, unlike health system leaders, many patients did not fully understand or trust the de-identification process. The platform's granular controls thus played a crucial role in providing reassurance: \textit{``[The platform] enables me to choose what information I'm willing to provide and ones I'm not. (P127)''}

\textbf{Patients Viewed Transparency as a Safeguard for the Outcomes of De-Identified Data Sharing. }Consistent with leader participants' views, patients also valued the platform's ability to provide transparency. Survey results showed that 415 (79.35\%) participants indicated it was important to know if their de-identified health information was being shared electronically, and 396 (75.72\%) expressed that knowing who had accessed their medical records was important. One participant stated: \textit{``If it [the platform] was available, it will help me in making decisions by knowing what my data will be used for and how it will be of benefit to me. (P154)''}

While leaders appreciated transparency as a way to inform patients about how and by whom their de-identified data would be used, patients often viewed transparency as a matter of accountability. Their concerns often extended beyond simply knowing who would access their data or what study it would support. Instead, patients saw transparency as a necessary safeguard to ensure that their de-identified data would be used ethically and lead to beneficial outcomes. Several patient participants expressed mistrust or concern about how the de-identified data might be used, especially in non-research or commercial contexts. P3 said: \textit{``Just this is who i am and i work in healthcare and know how beneficial data is. But i have my fears of using them against me as well (lack of trust maybe)''}. Even when the data was intended for research, some participants expressed concern about unethical use, such as \textit{``my medical condition being used for unethical experiments or research. (P332)''} Others were particularly worried about bad or discriminatory consequences if their data were misused: \textit{`` I am concerned that my medical record data may be used for discriminatory actions, such as insurance companies refusing to provide coverage or raising premiums. (P53) ''} For patients, transparency enabled by such a platform would serve not only as an informational tool but also as a safeguard against unethical or inappropriate data use. P86 said: \textit{``Most importantly, the digital consent process can ensure that my personal information is not stolen or misused, while also providing convenient access and management methods. ''}

\subsection{Health system leaders' and Patients' Perceived tensions and challenges with allowing patients to control which data is shared for secondary data research}


\subsubsection{Leaders' Perceived Challenges}

Interview data revealed three main tensions and challenges expressed by leader participants.



\textbf{Individual Preferences versus Unbiased Research Outcomes and Future Benefits.} Leader participants were concerned that patients' data-sharing autonomy might negatively impact research quality and the public health benefits if individuals were able to opt-out of their data for specific studies or clinical trials. L15 said: \textit{``I can see where there could be a concern that if you give people the option to opt out, then [it will] influence which records are available for research. (L15)''} Leader participants worried about the quality of research if many patients chose not to opt-in through patient-controlled platforms: \textit{``You're still limiting the data that you have available. You are potentially reducing the accuracy and validity of the results. (L6)''} 

Regarding patients' personal health benefits, leader participants worried that patients might focus solely on the risks of sharing data based on their current health conditions but may not be able to recognize the potential benefits of participating in research, leading them to make choices that may unexpectedly harm their future health. L13 said: \textit{``I also think that cutting people off from potential future treatments could be short-sighted or unethical''} since \textit{``we don't know what might happen in 2, 3, 4 years with their health, and so to exclude them from any potential opportunities in the future of a prospective clinical trial or research project that might benefit them if they have diabetes or a heart issue, or needs a tavern or cancer, you know, for those therapies that would be unethical. (L13)''} 


\textbf{Challenges of the Granular Consent System.} While leader participants perceived the design concept's granularity as a benefit, they also worried that this granularity might overwhelm or confuse some patients. Leader participants worried that the granularity might making it difficult for patients to make decisions when being presented with too many options. L4 doubted patients' ability to make granular decisions: \textit{`` I don't think it's appropriate for [us] to ask someone 'Oh, I want it to be used for this study, but not that study' because I don't think they have the background to be able to make those granular decisions. (L4)''} Leader participants felt it would be challenging for patients to figure out the nuanced meanings of the terms presented in the prototype, such as non-profit and for-profit institutions. L9 questioned the system's capability in explaining terms effectively: \begin{displayquote}\textcolor{black}{\textit{``It's not so simple right to click here and read what's for-profit… I'm worried about the fact that a large number of our patients would not be able to do this without the help of somebody. (L9)''}}\end{displayquote} 
Leader participants also worried that the system's granularity might increase patients' information overload, making them feel annoyed and even quit using the system when they had to review detailed information from any studies requesting their data. 
L9 was concerned that navigating the platform would demand considerable effort and manual guidance, which could not be done in a short period of time: \textit{``There's a lot of questions. It takes a lot of thinking. It takes a lot of time. It takes a lot of educate. This can't just be done in 5 min. Someone has to explain a lot. So it's complicated. It takes time. It's time consuming. (L9)''} 


\textbf{Negative Impact on Certain Patient Populations.} Leader participants worried certain patients, such as the elderly, non-native speakers, and individuals with low technology proficiency, might struggle with using a digital platform, designed to provide granular control, to decide their personal data-sharing choices. For instance, L9's institution served a significant number of patients aged 80 and above, worrying that these patients would have difficulty using the platform: \textit{``My own personal opinion of it is that it's too advanced. We have a number of our patients over 75 years old. All of these patients are not savvy with a computer. They're not going to be able to go on a computer and do this. (L9)''} 
L11 pointed out the practical challenges of translation, emphasizing a need to make the platform accessible in multiple languages while most existing medical materials were only available in English and Spanish: \textit{``What about people who can't read? What about people who don't read English or Spanish, because most things, you translate into Spanish. [But] There are so many people out there that don't read both of those. And it's just really hard. (L11)''} 

Leader participants also speculated that certain socioeconomic backgrounds or minoritized groups might be reluctant to share their data, and patient-controlled platforms would give them a chance to opt out and potentially skew research results that may potentially benefit them in the long run. L2 worried that allowing patients to selectively opt-in to research could create bias by only including those who completely undestand and feel satisfied with the the types of organizations requesting their data: \begin{displayquote}\textcolor{black}{\textit{``The biggest thing to me is that it creates a big biased dataset by only [including] those that said yes to government or whatever. Maybe they have a particular view, and maybe they have certain behaviors because of that view, right?  Societal views sometimes make them make a choice. (L2)''}}\end{displayquote} 
L6 highlighted concerns about research that utilized patients' data to train AI or machine learning algorithms, particularly if certain socioeconomic groups withheld their data due to a lack of trust in researchers or the government: \textit{``There are clear demographic trends in terms of people's trust in the government or medical researchers. And you would potentially be introducing bias into your datasets. If you have certain segments of the population that were removing their data and you're using that data, maybe to train AI algorithms or machine learning algorithms. You're introducing important bias into those algorithms. (L6)''}

\subsubsection{Patients' Perceived Challenges}

While patient participants generally valued the transparency and autonomy enabled by the prototype, many also raised concerns that aligned with or complicated the challenges identified by health system leaders. 

\textbf{Lack of Clear Personal Benefit Reduced Patient Motivation to Use the Platform. }While many patient participants recognized the societal benefits of sharing their de-identified data for research, some expressed a lack of motivation to use the platform due to unclear or insufficiently personalized benefits conveyed in the prototype. P8 said: \textit{``I can't think of anyone I know who would participate and if I told my children that I did they would not be too happy and give me plenty of reasons why.''} Another participant also expressed: \textit{``I may feel that this plan has not provided me with sufficient information or incentives, making me feel unwilling to participate. (P526)''} Some wanted access to the results of the studies that they would contribute to as a way to feel recognized and valued for their participation: \textit{`` I rarely am not open to sharing data. However, it is important to me that I also have an opportunity to receive the study results. (P4)''} Specifically, some participants expressed concerns about the effort required for long-term use of such a granular platform, noting that without clearly demonstrated benefits, patients would be reluctant to engage with the system over time. P528 said: \textit{``I may be skeptical of the long-term benefits of this plan, and I think it is not worth my time and energy.''} These views resonate with leaders' concerns that patients may prioritize concerns and burdens (e.g., the complexity of granular controls and privacy concerns) over potential long-term health benefits, especially when future benefits are hard to foresee.


\textbf{Patient Willingness Contrasted Leaders' Expectations: Those with Higher Health Literacy Were Less Likely to Use the Platform. }Survey results revealed that three groups of patients were more likely to disagree that they would personally use the platform: those with higher privacy concerns (odd ratio: 1.65, p-value<0.01), higher health literacy level (odd ratio: 1.36, p-value<0.01), and those who are older adults (odd ratio: 1.09,  p-value<0.01). The trends among patient participants with greater privacy concerns and older age were consistent with health system leaders' assumptions that such individuals might be more hesitant to adopt such as a granular data-sharing platform. However, the finding regarding highly health-literate participants conflicted with leaders' assumptions that patients with lower health literacy would be reluctant to engage with granular data-sharing tools. Qualitative results may explain such tendency from two aspects. First, some patient participants with high health literacy knew that de-identified data required no consent according to current regulations, therefore feeling no need for using an additional platform. For example, one patient participant said: \begin{displayquote}\textcolor{black}{\textit{``If it were to be used for secondary research on health data that would typically qualify for a waiver of consent and authorization, then I would not use it. With appropriate de-identification, data security and proper ethical and regulatory review the current regulatory framework (waivers of consent and HIPAA authorization) should suffice. (P19)''}}\end{displayquote} Second, patients with high health literacy may feel confident in their own ability to manage privacy independently, therefore no need for an extra platform. P521 said: \textit{``I may think that I can better manage and protect my medical information without using [the prototype].''}

\textbf{Patients Echoed Concerns About Health Inequity and Data Bias. }Aligned with leader participants' concerns, some patient participants also expressed worry about having such a platform might potentially exacerbate health inequity: \textit{``Allow selective opt-out, particularly via MyChart will further amplify health inequity in research and practice. (P19)''} Patient participants therefore shared a concern similar to that of leader participants regarding medical research: \textit{``Such a system would skew the participants in research, likely to the detriment of already under-represented populations. In addition, allowing for the exclusion of certain types of data would result in incomplete datasets. (P69)''}

\section{Discussion}

\textcolor{black}{Overall, our findings illustrate both stakeholders' support for a patient-controlled platform to manage de-identified data sharing, with both leader participants and patient participants valuing its potential to enhance patient autonomy through increased transparency and granular control. However, they also express important concerns about the potential burdens of granular decision-making, the risk of undermining research quality if patients selectively withhold data or limit participation in studies, and the possibility that such systems may exacerbate inequalities in participation. These findings surface critical tensions between individual control and societal benefit, between simplicity and informed, granular choice, and between empowerment and unintentional exclusion. We next delve into the major tensions identified in the study, unpacking the risks and benefits of granular consent, the differing perspectives by stakeholder groups, and the design implications that follow.}

\subsection{Understanding the risks and benefits of granular consent
}

Existing research highlights the importance of providing patients with granular control over their health data, especially in contexts of sharing for medical research. Studies show that patients often feel the current consent process lacks sufficient granularity, leaving them unable to control which specific types of data are shared. Sensitive data, such as mental health information, substance abuse history, domestic violence records, and sexual orientation, have been found to significantly influence individuals' willingness to share. Similarly, research in HCI reveals that even within intimate social networks, individuals are often more reluctant to share highly personal health information with certain stakeholders, such as parents or friends, than with partners or healthcare providers \cite{Lu2024b}. These findings emphasize the critical need for granularity in data-sharing platforms. That is, technology should allow individuals to manage their preferences according to the sensitivity of the data and their trust toward the data requester.

However, our study surfaces shared skepticism among both health system leaders and patients regarding the implementation of granularity in patient-controlled data-sharing platforms. Take leaders' concerns as an instance, their concerns mainly stem from the complexity and nuances of medical and research information, which they fear may exceed patients' literacy levels, thereby hindering their ability to make fully informed decisions. The complex nature of medical data, combined with the potential and invisible risks and benefits of sharing different types of information, poses a challenge for patients when faced with granular consent options. From the patient perspective, our findings show that many participants were concerned that such complexity would become overwhelming, particularly in the long run. Without clearly perceivable individual benefits, several participants expressed hesitation in putting the effort required to continuously manage data-sharing preferences, suggesting that the granularity intended to empower users could, paradoxically, deter engagement.

Stakeholders' such concerns and skepticism align with the privacy issues identified in existing tracking and health-related apps, particularly regarding uses' challenges with understanding app's privacy policies and user agreements with poor readability \cite{Alhajri2022, Robillard2019, Obar2020}. Additionally, one study found that users tend to skip reading apps' privacy policy and quickly accepting user agreement due to the experienced information overload \cite{Obar2020}. This behavior may stem from users' perceiving certain personal health data, such as diet, exercises, and sleep, as having little value or sensitivity \cite{Ajana2020}. However, medical data usually contains sensitive information such as mental health records and substance abuse history, making patients hesitant to share these details. Therefore, the information overload associated with the difficulty of handling granularity, combined with the sensitivity of medical data, may greatly confuse patients and lead to more rejections. \textcolor{black}{Work in usable privacy controls further shows that more granular choices do not automatically translate into better control. Proliferating privacy options can impose high cognitive and time burden and disrupt routine use, which may actually reduce meaningful control because of fatigue, confusion, and negative emotional consequences \cite{Zimmeck2024GeneralizableAP,feng2021designspace,korff2014toomuch}}.

Overall, granular consent presents a \textcolor{black}{fundamental} trade-off between empowering patients and maintaining \textcolor{black}{usability and} comprehensive care. Health leaders appreciate the promise of greater patient agency, but remain cautious about the risks to care continuity and added complexity in workflows. \textcolor{black}{Patients welcome the opportunity to safeguard sensitive information and tailor sharing, yet worry about potential misuse and the burden of making too many decisions. Taken together, our findings suggest that a successful patient-controlled data-sharing platform must balance the flexibility brought by granularity with simplicity through interfaces and mechanisms that minimize unnecessary interactions while still supporting meaningful consent decisions.}


\subsection{Divergent Perceptions of a Patient-Controlled Data Platform}

Our study reveals that health system leaders and patients share common values of transparency and choice in a patient-controlled data-sharing platform, but their underlying motivations and concerns differ in various ways. 

Health system leaders approach the platform with a focus on institutional ethics and operational feasibility. Their emphasis lies in informed consent, patient autonomy, and advancing the greater good. In contrast, patients, while supportive of these same features, tend to interpret them through the lens of outcome risks, privacy, and personal benefits. They embrace features that support their agency, such as transparency and granular control, not only as ethical ideals but also as practical tools for self-protection and personal reward. These differing orientations could create misalignments in the design and implementation of patient-controlled platforms for de-identified data sharing. For instance, leaders might believe transparency is achieved by just providing a clear consent form, whereas a patient may only feel secure if transparency includes ongoing updates or reciprocal benefits, such as receiving study results. Similarly, leaders may view granular controls as participation incentives compared to the traditional \textit{``All-or-Nothing''} approach, while patients may see them as essential safeguards for navigating the emotional and ethical uncertainties of sharing sensitive health information. 

Our findings also highlight the critical role of trust and knowledge in technology adoption: providing users with control alone does not ensure engagement,\textbf{ }especially when more health-literate or tech-savvy patients remain unconvinced of a platform's trustworthiness or necessity\textbf{. }Contrary to common assumptions that individuals with lower health literacy or educational level are more likely to avoid new technologies \cite{mackert2016health, manganello2017relationship,uematsu2010can}, our survey revealed the opposite trend. In this paper, patients with higher health literacy were actually less likely to use the platform. This seemingly paradoxical result suggests that greater knowledge may reduce technology adoption, as heightened confidence may diminish the perceived need for external privacy tools or controls. 
These findings challenge oversimplified design assumptions that equate ``low literacy'' with ``low engagement'' and emphasize the need for more nuanced approaches when designing patient-facing data-sharing technologies. 

The tension between individual control and research needs is recognized by both stakeholders, realizing that empowering patients must be balanced with sustaining data integrity and equity. Prior work has highlighted that data sharing for the public good cannot succeed without public trust and involvement \cite{Jamieson2021,Lu2022}. Our study adds nuance by revealing that patients build trust through transparency and reassurance against potential risks, while leaders emphasize trust through formal consent procedures and the principle of respecting autonomy. \textcolor{black}{In the later design implication section, we propose ways to foster patient engagement with the system, helping ensure sufficient data contributions for research quality}. 

By critically examining the gaps between leaders' and patients' viewpoints, this work underscores that empowering patients with data control is not merely a technical design task, but a socio-technical balancing act between agency and altruism as well as privacy and benefits. 

\subsection{Design Implications}

Based on prior literature and insights obtained from this study, we propose a flexible, context-aware approach to granular consent, one that not only allows users to tailor data-sharing decisions based on data sensitivity and trust but also emphasizes personal benefit and adapts to users' varying levels of literacy.



\textcolor{black}{ \textbf{Supporting Flexible\textcolor{black} { , Low-burden Granular Data Sharing: We recognize not all data requests necessarily require granularity.}} Data-sharing platforms should enable flexible levels of granularity, allowing patients to tailor their sharing preferences according to the perceived sensitivity of the de-identified data and their trust in the recipient organization. A tiered structure of granularity can be presented to patients. For example, if a patient does not perceive a high-level data category as sensitive, granular sharing options for subtypes under this category may not be shown to patients for review. To help patients make informed decisions, the systems should remind patients when highly sensitive data are involved, such as the pregnancy-related information identified in our study. Confirmation popups or visual aids, such as highlighting or color coding, could be utilized to alert patients that they are dealing with sensitive data. \textcolor{black}{At the same time, to address concerns about the potential burdens introduced by fine-grained controls, the platform should incorporate features that streamline and augment the granular consent process through reducing repetitive effort and supporting ``configure once, reuse often'' patterns, as detailed below. For instance, the system should allow patients to whitelist certain trusted institutions or designate less sensitive data types to reduce repetitive effort, aligning with prior work that advocates a generalizable setup across contexts and reserves individual decisions only when necessary\cite{Zimmeck2024GeneralizableAP}. Building on this, we can imagine a conversational feature to enhance the current default preference configuration in the system. Instead of asking patients to manually complete a static matrix of choices, the system could conduct a brief, guided survey to elicit their attitudes and preferences about sharing different categories of data with different types of recipients. From this interaction, it could infer an initial preference profile, automatically apply it to incoming requests, and still allow patients to override or refine specific decisions over time. When multiple requests or data types are being reviewed, the system should support bulk actions to help patients efficiently apply consistent preferences across similar items. Further reducing cognitive burden, a smart and adaptive system can provide automated and personalized recommendations \cite{feng2021designspace,Smullen2020BestOfBothWorlds} for sharing preferences based on consent history when patients receive new requests with similar characteristics. Patients could accept the system’s recommendation in one step or expand to make granular choices, which helps minimize cognitive burden while still preserving control on demand. Finally, we acknowledge that these strategies may not fully eliminate the trade-off between granular control and user burden. Acquiring more fine-grained control over data sharing will likely always require some effort from patients to articulate and revise their preferences. Rather than assuming that most users will meticulously configure every setting, the design goal is to ensure that those who \textit{do} wish to exercise granular control can do so with minimal avoidable burden, while others can rely on well-calibrated, transparent defaults. While the tension between granularity and user burden is unlikely to disappear, designers should work on tools can help ensure that this tension is acknowledged, managed, and distributed in ways that respect patient autonomy without overwhelming them.}} 


\textcolor{black}{
\textbf{Empowering Patients Through Ongoing, Benefit-Centered Transparency.}
To address the tension between patients' data-sharing autonomy and the broader societal benefits of research, de-identified data-sharing platforms should prioritize communicating the potential benefits of data sharing, especially on individual levels. Insights from previous studies suggest that patients tend to be more willing to share their data when they can clearly perceive the benefits \cite{Benevento2023}. Medical research indicates that patients often share their data out of altruism or to enhance their own medical knowledge and get access to the treatments provided in the research project \cite{Morse2023,Kalkman2019}. These findings all highlight that when individual benefits are clearly articulated, patients are more motivated to share data and are less likely to be deterred by concerns such as privacy. Patients may also find it difficult to foresee the long-term advantages of sharing their data, it is therefore crucial to present these benefits in a way that resonates with their personal health goals. For instance, platforms might include dashboards or personalized summaries showing how a patient's de-identified data contributed to clinical trials or research studies, especially those focusing on conditions relevant to the patient. Patients could be notified if their data helped identify patterns or findings leading to the development of new medications or treatments, or if they become eligible to be contacted for innovative clinical trials, even for conditions they do not currently have. This could help patients recognize how sharing data today may benefit them in the future through gaining early access to research opportunities or medical interventions. In addition, our study surfaces nuanced differences in how transparency is perceived: while leaders see it as informing patients about the use of their data during the consent process, patients expect ongoing updates about how their data are used and the outcomes of their participation. This suggests that transparency in such systems should be designed as an interactive and continuous experience, instead of a one-time disclosure. In particular, data-sharing consent should move beyond standardized forms that merely list who is requesting the data and for what purpose, and instead providing patients with insight into the progress and outcome of studies that used their data. A progress tracker or map could display how their contributions advanced medical research (e.g., publications) and how many patients have benefited from previous data-sharing of similar nature. 
Additionally, systems could present aggregated statistics and easy-to-grasp figures to help individuals contextualize their decisions and understand peer contributions. }

\textcolor{black}{\textbf{Addressing Nuanced Privacy Perspectives and Data Engagement Needs.}
Contrary to leader participants' perceptions that patients with low health literacy may be reluctant to engage with data-sharing platforms, our findings reveal a more complex dynamic. That is, patients with high health literacy, in fact, may be more likely to disengage, not due to confusion or privacy concerns, but because of their confidence in the safety of de-identified data and their own ability to manage privacy independently. This suggests 
that, rather than adopting a one-size-fits-all approach, researchers or designers must recognize that different literacy profiles may lead to disengagement for completely different reasons, either confusion and overwhelm or overconfidence and dismissal. Therefore, the information required for data-sharing consent should be adapted to match patients' varying levels of health and data literacy, ensuring that all users, regardless of background, can make informed decisions. 
}
\textcolor{black}{Rather than creating entirely separate pathways for ``low'' versus ``high'' literacy users, platforms should offer layered, just-in-time support that is available to everyone and avoid hard-coding assumptions about users' literacy from static demographic cues (e.g., education and profession). For example, baseline explanations of data categories and data uses, as well as the potential risks and benefits of sharing, should be accessible to all users in clear, plain language with intuitive examples and visual cues, reflecting design principles such as concision, transparency, and valuing participants' time and effort \cite{McInnis2024explore}, so that critical protections do not require extra work from any particular group. For those seeking more detail, additional layers of explanation can be offered as optional expansions
, with different literacy profiles leading to \textit{different} kinds of work (e.g., clarifying complex concepts versus revisiting overconfident assumptions) rather than simply more or less effort. An AI-powered conversational agent could serve as a flexible support tool by 
proactively surfacing information based on user behavior. 
Critically, such an agent should not assume user literacy based on basic demographic profiles or group patients into broad, institution-defined literacy levels \cite{liu2020meaning,MalloyWeir2016,Urstade056294}. Instead, the system should infer informational needs from actual interaction patterns (e.g., hesitation, repeated help requests, or quickly skipping key explanations) and then adaptively adjust the amount, format, and timing of support. For example, if a user hesitates over a medical term or repeatedly asks about risks, the system could offer progressively detailed answers, visual aids, or even audio alternatives \cite{glaser2020interventions}, or tabular summaries\cite{Brick2020Risk} to help bridge understanding. Conversely, for users who appear highly confident and quickly dismiss risk information, the agent might surface targeted explanations about re-identification and governance to gently challenge assumptions that de-identified data are ``always safe''. This approach acknowledges that users with higher literacy can still misinterpret basic concepts, while users with lower literacy may seek deeper information but feel discouraged by complex medical terminology. Ultimately, systems should ensure that all users, regardless of background, are empowered to make informed, autonomous decisions, without structurally imposing greater effort on those already disadvantaged by gaps in health communication or technology access \cite{Breese2007evaluation}.}



\section{Conclusion}

This study advances HCI and CSCW research by providing a comparative investigation of how patients and health system leaders perceive a granular, patient-controlled data-sharing platform for de-identified medical data. By combining qualitative insights from institutional decision-makers with quantitative and qualitative data from a diverse sample of patients, we surface both alignment and tension in stakeholder priorities. While both groups support greater transparency and control, leaders emphasize ethical compliance and research continuity, whereas patients seek protection, accountability, and meaningful personal benefits. These differences reveal that empowering patients requires more than offering choice. It demands systems that are responsive to diverse personal values, privacy concerns, and literacy levels. As patient-controlled data-sharing platforms become more common, especially in secondary research settings, our study highlights the importance of designing for flexibility, clarity, and reciprocity.

\begin{acks}
We thank our anonymous reviewers for their feedback. This work was supported in part by NIH/NHGRI under Award R01HG011066.
\end{acks}

\bibliographystyle{ACM-Reference-Format}
\bibliography{paper}

@article{Smullen2020BestOfBothWorlds,
  title   = {The Best of Both Worlds: Mitigating Trade-offs Between Accuracy and User Burden in Capturing Mobile App Privacy Preferences},
  author  = {Smullen, Daniel and Feng, Yuanyuan and Zhang, Shikun (Aerin) and Sadeh, Norman},
  journal = {Proceedings on Privacy Enhancing Technologies},
  year    = {2020},
  volume  = {2020},
  number  = {1},
  pages   = {195--215},
  doi     = {10.2478/popets-2020-0011}
}

@article{MalloyWeir2016,
  author  = {Malloy-Weir, Leslie J and Charles, Cathy and Gafni, Amiram and Entwistle, Vikki},
  title   = {A review of health literacy: Definitions, interpretations, and implications for policy initiatives},
  journal = {Journal of Public Health Policy},
  year    = {2016},
  volume  = {37},
  number  = {3},
  pages   = {334--352},
  doi     = {10.1057/jphp.2016.18},
  url     = {https://doi.org/10.1057/jphp.2016.18}
}

@article {Urstade056294,
	author = {Urstad, Kristin Hjorthaug and Andersen, Marit Helen and Larsen, Marie Hamilton and Borge, Christine R{\r a}heim and Helseth, S{\o}lvi and Wahl, Astrid Klopstad},
	title = {Definitions and measurement of health literacy in health and medicine research: a systematic review},
	volume = {12},
	number = {2},
	elocation-id = {e056294},
	year = {2022},
	doi = {10.1136/bmjopen-2021-056294},
	publisher = {British Medical Journal Publishing Group},
	issn = {2044-6055},
	URL = {https://bmjopen.bmj.com/content/12/2/e056294},
	journal = {BMJ Open}
}

@article{liu2020meaning,
  title   = {What is the meaning of health literacy? A systematic review and qualitative synthesis},
  author  = {Liu, Chenxi and Wang, Dan and Liu, Chaojie and Jiang, Junnan and Wang, Xuemei and Chen, Haihong and Ju, Xin and Zhang, Xinping},
  journal = {Family Medicine and Community Health},
  year    = {2020},
  volume  = {8},
  number  = {2},
  pages   = {e000351},
  doi     = {10.1136/fmch-2020-000351},
  url     = {https://doi.org/10.1136/fmch-2020-000351}
}

@inproceedings {korff2014toomuch,
author = {Stefan Korff and Rainer B{\"o}hme},
title = {Too Much Choice: {End-User} Privacy Decisions in the Context of Choice Proliferation},
booktitle = {10th Symposium On Usable Privacy and Security (SOUPS 2014)},
year = {2014},
isbn = {978-1-931971-13-3},
address = {Menlo Park, CA},
pages = {69--87},
url = {https://www.usenix.org/conference/soups2014/proceedings/presentation/korff},
publisher = {USENIX Association},
month = jul
}

@inproceedings{feng2021designspace,
author = {Feng, Yuanyuan and Yao, Yaxing and Sadeh, Norman},
title = {A Design Space for Privacy Choices: Towards Meaningful Privacy Control in the Internet of Things},
year = {2021},
isbn = {9781450380966},
publisher = {Association for Computing Machinery},
address = {New York, NY, USA},
url = {https://doi.org/10.1145/3411764.3445148},
doi = {10.1145/3411764.3445148},
booktitle = {Proceedings of the 2021 CHI Conference on Human Factors in Computing Systems},
articleno = {64},
numpages = {16},
location = {Yokohama, Japan},
series = {CHI '21}
}

@article{Zimmeck2024GeneralizableAP,
  title={Generalizable Active Privacy Choice: Designing a Graphical User Interface for Global Privacy Control},
  author={Sebastian Zimmeck and Eliza Dahlia Kuller and Chunyue Ma and Bella Tassone and Joe Champeau},
  journal={Proc. Priv. Enhancing Technol.},
  year={2024},
  volume={2024},
  pages={258-279},
  doi={https://doi.org/10.56553/popets-2024-0015}
}

@article{Breese2007evaluation,
author = {Peter E. Breese and William J. Burman and Stefan Goldberg and Stephen E. Weis},
title = {Education Level, Primary Language, and Comprehension of the Informed Consent Process},
journal = {Journal of Empirical Research on Human Research Ethics},
year = {2007},
doi = {10.1525/jer.2007.2.4.69},
    note ={PMID: 19385809},
URL = { 
    
        https://doi.org/10.1525/jer.2007.2.4.69
},
eprint = { 
    
        https://doi.org/10.1525/jer.2007.2.4.69 

}
}

@article{Brick2020Risk,
  title   = {Risk communication in tables versus text: a registered report randomized trial on 'fact boxes'},
  author  = {Brick, Cameron and McDowell, Michelle and Freeman, Alexandra L. J.},
  journal = {Royal Society Open Science},
  volume  = {7},
  number  = {3},
  pages   = {190876},
  year    = {2020},
  doi     = {10.1098/rsos.190876}
}

@article{McInnis2024explore,
author = {McInnis, Brian James and Pindus, Ramona and Kareem, Daniah and Gamboa, Savannah and Nebeker, Camille},
title = {Exploring the Future of Informed Consent: Applying a Service Design Approach},
year = {2024},
issue_date = {April 2024},
publisher = {Association for Computing Machinery},
address = {New York, NY, USA},
volume = {8},
number = {CSCW1},
url = {https://doi.org/10.1145/3637330},
doi = {10.1145/3637330},
journal = {Proc. ACM Hum.-Comput. Interact.},
month = apr,
articleno = {53},
numpages = {31}
}

@article{glaser2020interventions,
author = {Johanna Glaser and Sarah Nouri and Alicia Fernandez and Rebecca L. Sudore and Dean Schillinger and Michele Klein-Fedyshin and Yael Schenker},
title ={Interventions to Improve Patient Comprehension in Informed Consent for Medical and Surgical Procedures: An Updated Systematic Review},

journal = {Medical Decision Making},
volume = {40},
number = {2},
pages = {119-143},
year = {2020},
doi = {https://doi.org/10.1177/0272989X19896348}
}

@inproceedings{uematsu2010can,
  title={Can education be a barrier to technology adoption?},
  author={Uematsu, Hiroki and Mishra, Ashok K},
  booktitle={2010 Annual Meeting, July 25-27, 2010, Denver, Colorado},
  number={61630},
  year={2010},
  organization={Agricultural and Applied Economics Association}
}

@article{mackert2016health,
  title={Health Literacy and Health Information Technology Adoption: the Potential for a New Digital Divide},
  author={Mackert, Michael and Mabry-Flynn, Amanda and Champlin, Sara and Donovan, Erin E and Pounders, Kathrynn},
  journal={Journal of medical Internet research},
  volume={18},
  number={10},
  pages={e264},
  year={2016},
  doi={10.2196/jmir.6349},
  publisher={JMIR Publications Toronto, Canada}
}

@article{manganello2017relationship,
  title={The Relationship of Health Literacy with Use of Digital Technology for Health Information: Implications for Public Health Practice},
  author={Manganello, Jennifer and Gerstner, Gena and Pergolino, Kristen and Graham, Yvonne and Falisi, Angela and Strogatz, David},
  journal={Journal of public health management and practice},
  volume={23},
  number={4},
  pages={380--387},
  year={2017},
  doi={10.1097/PHH.0000000000000366},
  publisher={LWW}
}

@article{parker1995test,
  title={The Test of Functional Health Literacy in Adults: a New Instrument for Measuring Patients’ Literacy Skills},
  author={Parker, Ruth M and Baker, David W and Williams, Mark V and Nurss, Joanne R},
  journal={Journal of general internal medicine},
  volume={10},
  pages={537--541},
  year={1995},
  doi={10.1007/BF02640361},
  publisher={Springer}
}

@article{nelson2004health,
  title={The Health Information National Trends Survey (HINTS): Development, Design, and Dissemination},
  author={Nelson, David and Kreps, Gary and Hesse, Bradford and Croyle, Robert and Willis, Gordon and Arora, Neeraj and Rimer, Barbara and Vish Viswanath, K and Weinstein, Neil and Alden, Sara},
  journal={Journal of health communication},
  volume={9},
  number={5},
  pages={443--460},
  year={2004},
  doi={10.1080/10810730490504233},
  publisher={Taylor \\& Francis}
}

@article{hesse2005trust,
  title={Trust and Sources of Health Information: the Impact of the Internet and its Implications for Health Care Providers: Findings from the First Health Information National Trends Survey},
  author={Hesse, Bradford W and Nelson, David E and Kreps, Gary L and Croyle, Robert T and Arora, Neeraj K and Rimer, Barbara K and Viswanath, Kasisomayajula},
  journal={Archives of internal medicine},
  volume={165},
  number={22},
  pages={2618--2624},
  year={2005},
  doi={10.1001/archinte.165.22.2618},
  publisher={American Medical Association}
}

@misc{USDepartmentofHealthandHumanServices2025,
   author = {US Department of Health and Human Services},
   month = {3},
   title = {Summary of the HIPAA Privacy Rule | HHS.gov},
   url = {https://www.hhs.gov/hipaa/for-professionals/privacy/laws-regulations/index.html},
   year = {2025}
}

@article{abdelhamid2017putting,
  title={Putting the Focus Back on the Patient: How Privacy Concerns Affect Personal Health Information Sharing Intentions},
  author={Abdelhamid, Mohamed and Gaia, Joana and Sanders, G Lawrence},
  journal={Journal of medical Internet research},
  volume={19},
  number={9},
  pages={e169},
  year={2017},
  publisher={JMIR Publications Toronto, Canada}
}

@article{platt2015public,
  title={Public Trust in Health Information Sharing: Implications for Biobanking and Electronic Health Record Systems},
  author={Platt, Jodyn and Kardia, Sharon},
  journal={Journal of personalized medicine},
  volume={5},
  number={1},
  pages={3--21},
  year={2015},
  publisher={MDPI}
}

@article{mello2018clinical,
  title={Clinical Trial Participants’ Views of the Risks and Benefits of Data Sharing},
  author={Mello, Michelle M and Lieou, Van and Goodman, Steven N},
  journal={New England journal of medicine},
  volume={378},
  number={23},
  pages={2202--2211},
  year={2018},
  doi={10.1056/NEJMsa1713258},
  publisher={Mass Medical Soc}
}

@article{perera2011views,
  title={Views on Health Information Sharing and Privacy from Primary Care Practices Using Electronic Medical Records},
  author={Perera, Gihan and Holbrook, Anne and Thabane, Lehana and Foster, Gary and Willison, Donald J},
  journal={International journal of medical informatics},
  volume={80},
  number={2},
  pages={94--101},
  year={2011},
  doi={10.1016/j.ijmedinf.2010.11.005},
  publisher={Elsevier}
}

@article{goodman2017identified,
  title={De-Identified Genomic Data Sharing: the Research Participant Perspective},
  author={Goodman, Deborah and Johnson, Catherine O and Bowen, Deborah and Smith, Megan and Wenzel, Lari and Edwards, Karen},
  journal={Journal of community genetics},
  volume={8},
  pages={173--181},
  year={2017},
  doi={10.1007/s12687-017-0300-1},
  publisher={Springer}
}

@article{whiddett2006patients,
  title={Patients’ Attitudes towards Sharing their Health Information},
  author={Whiddett, Richard and Hunter, Inga and Engelbrecht, Judith and Handy, Jocelyn},
  journal={International journal of medical informatics},
  volume={75},
  number={7},
  pages={530--541},
  year={2006},
  doi={10.1016/j.ijmedinf.2005.08.009},
  publisher={Elsevier}
}

@article{goodman2018comparison,
  title={A Comparison of Views Regarding the Use of De-Identified Data},
  author={Goodman, Deborah and Johnson, Catherine O and Bowen, Deborah and Smith, Megan and Wenzel, Lari and Edwards, Karen L},
  journal={Translational Behavioral Medicine},
  volume={8},
  number={1},
  pages={113--118},
  year={2018},
  doi={10.1093/tbm/ibx054},
  publisher={Oxford University Press US}
}

@article{el2011systematic,
  title={A Systematic Review of Re-Identification Attacks on Health Data},
  author={El Emam, Khaled and Jonker, Elizabeth and Arbuckle, Luk and Malin, Bradley},
  journal={PloS one},
  volume={6},
  number={12},
  pages={e28071},
  year={2011},
  doi={10.1371/journal.pone.0028071},
  publisher={Public Library of Science San Francisco, USA}
}

@article{murdoch2021privacy,
  title={Privacy and Artificial Intelligence: Challenges for Protecting Health Information in a New Era},
  author={Murdoch, Blake},
  journal={BMC medical ethics},
  volume={22},
  pages={1--5},
  year={2021},
  doi={10.1186/s12910-021-00687-3},
  publisher={Springer}
}

@article{packhauser2022deep,
  title={Deep Learning-Based Patient Re-Identification is Able to Exploit the Biometric Nature of Medical Chest X-Ray Data},
  author={Packh{\"a}user, Kai and G{\"u}ndel, Sebastian and M{\"u}nster, Nicolas and Syben, Christopher and Christlein, Vincent and Maier, Andreas},
  journal={Scientific Reports},
  volume={12},
  number={1},
  pages={14851},
  year={2022},
  doi={https://doi.org/10.1038/s41598-022-19045-3},
  publisher={Nature Publishing Group UK London}
}

@article{trinidad2020public,
  title={The Public’s Comfort with Sharing Health Data with Third-Party Commercial Companies},
  author={Trinidad, M Grace and Platt, Jodyn and Kardia, Sharon LR},
  journal={Humanities and Social Sciences Communications},
  volume={7},
  number={1},
  pages={1--10},
  year={2020},
  doi={https://doi.org/10.1057/s41599-020-00641-5},
  publisher={Palgrave}
}

@article{chiruvella2021ethical,
  title={Ethical Issues in Patient Data Ownership},
  author={Chiruvella, Varsha and Guddati, Achuta Kumar and others},
  journal={Interactive journal of medical research},
  volume={10},
  number={2},
  pages={e22269},
  year={2021},
  doi={10.2196/22269},
  publisher={JMIR Publications Inc., Toronto, Canada}
}

@article{Clemens2012,
   author = {Norman A. Clemens},
   doi = {10.1097/01.PRA.0000410987.38723.47},
   issn = {15274160},
   issue = {1},
   journal = {Journal of Psychiatric Practice},
   month = {1},
   pages = {46-50},
   pmid = {22261983},
   title = {Privacy, Consent, and the Electronic Mental Health Record: The Person vs. the System},
   volume = {18},
   url = {https://journals.lww.com/practicalpsychiatry/fulltext/2012/01000/privacy,_consent,_and_the_electronic_mental_health.7.aspx},
   year = {2012},
}

@article{Caulfield2007,
   author = {Timothy Caulfield},
   doi = {10.1080/09615768.2007.11427674},
   issn = {0961-5768},
   issue = {2},
   journal = {King's Law Journal},
   month = {1},
   pages = {209-226},
   publisher = {Routledge},
   title = {Biobanks and Blanket Consent: The Proper Place of the Public Good and Public Perception Rationales},
   volume = {18},
   url = {https://www.tandfonline.com/doi/abs/10.1080/09615768.2007.11427674},
   year = {2007},
}

@article{Grekousis2021,
   author = {George Grekousis and Ye Liu},
   doi = {10.1016/J.SCS.2021.102995},
   issn = {2210-6707},
   journal = {Sustainable Cities and Society},
   month = {8},
   pages = {102995},
   publisher = {Elsevier},
   title = {Digital Contact Tracing, Community Uptake, and Proximity Awareness Technology to Fight COVID-19: a Systematic Review},
   volume = {71},
   year = {2021},
}

@article{Zimmermann2021,
   author = {Bettina Maria Zimmermann and Amelia Fiske and Barbara Prainsack and Nora Hangel and Stuart McLennan and Alena Buyx},
   doi = {10.2196/25525},
   issue = {2},
   journal = {Journal of Medical Internet Research},
   month = {2},
   pages = {e25525},
   publisher = {Journal of Medical Internet Research},
   title = {Early Perceptions of COVID-19 Contact Tracing Apps in German-Speaking Countries: Comparative Mixed Methods Study},
   volume = {23},
   year = {2021},
}

@article{Altmann2020,
   author = {Samuel Altmann and Luke Milsom and Hannah Zillessen and Raffaele Blasone and Frederic Gerdon and Ruben Bach and Frauke Kreuter and Daniele Nosenzo and Séverine Toussaert and Johannes Abeler},
   doi = {10.2196/19857},
   issue = {8},
   journal = {JMIR mHealth and uHealth},
   month = {8},
   pages = {e19857},
   publisher = {JMIR mHealth and uHealth},
   title = {Acceptability of App-Based Contact Tracing for COVID-19: Cross-Country Survey Study},
   volume = {8},
   year = {2020},
}

@article{Ajana2020,
   author = {Btihaj Ajana},
   doi = {https://doi.org/10.1177/0539018420959522},
   issn = {14617412},
   issue = {4},
   journal = {Social Science Information},
   month = {12},
   pages = {654-678},
   publisher = {SAGE Publications Ltd},
   title = {Personal Metrics: Users’ Experiences and Perceptions of Self-Tracking Practices and Data},
   volume = {59},
   url = {https://journals.sagepub.com/doi/full/10.1177/0539018420959522},
   year = {2020},
}

@inproceedings{Mehrnezhad2020,
   author = {Maryam Mehrnezhad},
   doi = {10.1109/EUROSPW51379.2020.00023},
   isbn = {9781728185972},
   booktitle = {IEEE European Symposium on Security and Privacy Workshops (EuroS\&PW 2020)},
   month = {9},
   pages = {97-106},
   publisher = {Institute of Electrical and Electronics Engineers Inc.},
   title = {A Cross-Platform Evaluation of Privacy Notices and Tracking Practices},
   year = {2020},
}

@inproceedings{Nguyen2022,
   author = {Trung Tin Nguyen and Michael Backes and Ben Stock},
   doi = {10.1145/3548606.3560564},
   isbn = {9781450394505},
   issn = {15437221},
   booktitle = {Proceedings of the ACM Conference on Computer and Communications Security (CCS 2022)},
   month = {11},
   pages = {2369-2383},
   publisher = {Association for Computing Machinery},
   title = {Freely Given Consent?: Studying Consent Notice of Third-Party Tracking and Its Violations of GDPR in Android Apps},
   url = {https://dl.acm.org/doi/10.1145/3548606.3560564},
   year = {2022},
}

@article{Hutton2018,
   author = {Luke Hutton and Blaine A Price and Ryan Kelly and Ciaran McCormick and Arosha K Bandara and Tally Hatzakis and Maureen Meadows and Bashar Nuseibeh},
   doi = {10.2196/mhealth.9217},
   issn = {2291-5222},
   issue = {10},
   journal = {JMIR mHealth and uHealth},
   month = {10},
   pages = {e185},
   pmid = {30348623},
   publisher = {JMIR mHealth and uHealth},
   title = {Assessing the Privacy of mHealth Apps for Self-Tracking: Heuristic Evaluation Approach.},
   volume = {6},
   url = {http://www.ncbi.nlm.nih.gov/pubmed/30348623 http://www.pubmedcentral.nih.gov/articlerender.fcgi?artid=PMC6231850},
   year = {2018},
}

@article{Obar2020,
   author = {Jonathan A. Obar and Anne Oeldorf-Hirsch},
   doi = {10.1080/1369118X.2018.1486870},
   issn = {14684462},
   issue = {1},
   journal = {Information, Communication \& Society},
   month = {1},
   pages = {128-147},
   publisher = {Routledge},
   title = {The Biggest Lie on the Internet: Ignoring the Privacy Policies and Terms of Service Policies of Social Networking Services},
   volume = {23},
   url = {https://www.tandfonline.com/doi/abs/10.1080/1369118X.2018.1486870},
   year = {2020},
}

@article{Robillard2019,
   author = {Julie M. Robillard and Tanya L. Feng and Arlo B. Sporn and Jen Ai Lai and Cody Lo and Monica Ta and Roland Nadler},
   doi = {10.1016/J.INVENT.2019.100243},
   issn = {2214-7829},
   journal = {Internet Interventions},
   month = {9},
   pages = {100243},
   publisher = {Elsevier},
   title = {Availability, Readability, and Content of Privacy Policies and Terms of Agreements of Mental Health Apps},
   volume = {17},
   year = {2019},
}

@article{Alhajri2022,
   author = {May Alhajri and Ahmad Salehi Shahraki and Carsten Rudolph},
   doi = {10.1145/3511616.3513100},
   isbn = {9781450396066},
   journal = {Proceedings of the 2022 Australasian Computer Science Week (ACSW 2022)},
   month = {2},
   pages = {65-73},
   publisher = {Association for Computing Machinery},
   title = {Privacy of Fitness Applications and Consent Management in Blockchain},
   url = {https://dl.acm.org/doi/10.1145/3511616.3513100},
   year = {2022},
}

@article{Oh2022,
   author = {Chi Young Oh and Yuhan Luo and Beth St. Jean and Eun Kyoung Choe},
   doi = {https://doi.org/10.1145/3512954},
   issn = {25730142},
   issue = {CSCW1},
   journal = {Proceedings of the ACM on Human-Computer Interaction},
   month = {4},
   publisher = {Association for Computing Machinery},
   title = {Patients Waiting for Cues: Information Asymmetries and Challenges in Sharing Patient-Generated Data in the Clinic},
   volume = {6},
   year = {2022},
}

@inproceedings{West2018,
   author = {Peter West and Max Van Kleek and Richard Giordano and Mark J. Weal and Nigel Shadbolt},
   doi = {https://dl.acm.org/doi/10.1145/3173574.3174058},
   isbn = {9781450356206},
   booktitle = {Proceedings of the SIGCHI Conference on Human Factors in Computing Systems (CHI 2018)},
   month = {4},
   title = {Common Barriers to the Use of Patient-Generated Data Across Clinical Settings},
   volume = {2018-April},
   year = {2018},
}

@article{Jamieson2021,
   author = {Jack Jamieson and Naomi Yamashita and Daniel A. Epstein and Yunan Chen},
   doi = {10.1145/3479868},
   issue = {2},
   journal = {Proceedings of the ACM on Human-Computer Interaction},
   pages = {30},
   title = {Deciding If and How to Use a COVID-19 Contact Tracing App: Influences of Social Factors on Individual Use in Japan},
   volume = {5},
   year = {2021},
}

@article{Seberger2021,
   author = {John S. Seberger and Sameer Patil},
   doi = {10.2196/30871},
   issn = {22915222},
   issue = {10},
   journal = {JMIR Mhealth Uhealth},
   month = {10},
   pages = {e30871},
   pmid = {34519667},
   publisher = {JMIR mHealth and uHealth},
   title = {Post-COVID Public Health Surveillance and Privacy Expectations in the United States: Scenario-Based Interview Study},
   volume = {9},
   year = {2021},
}

@inproceedings{Utz2021,
   author = {Christine Utz and Stefen Becker and Theodor Schnitzler and Florian M. Farke and Franziska Herbert and Leonie Schaewitz and Martin Degeling and Markus Dürmuth and Markus Dür},
   city = {New York, NY, USA},
   doi = {10.1145/3411764.3445517},
   journal = {Proceedings of the SIGCHI Conference on Human Factors in Computing Systems (CHI 2021)},
   title = {Apps Against the Spread: Privacy Implications and User Acceptance of COVID-19-Related Smartphone Apps on Three Continents},
   volume = {22},
   year = {2021},
}

@article{Freifeld2010,
   author = {Clark C. Freifeld and Rumi Chunara and Sumiko R. Mekaru and Emily H. Chan and Taha Kass-Hout and Anahi Ayala Iacucci and John S. Brownstein},
   doi = {10.1371/journal.pmed.1000376},
   issn = {1549-1676},
   issue = {12},
   journal = {PLoS Medicine},
   month = {12},
   pages = {e1000376},
   publisher = {Public Library of Science},
   title = {Participatory Epidemiology: Use of Mobile Phones for Community-Based Health Reporting},
   volume = {7},
   url = {https://dx.plos.org/10.1371/journal.pmed.1000376},
   year = {2010},
}

@inproceedings{Tadas2023,
   author = {Shreya Tadas and Jane Dickson and David Coyle},
   doi = {https://dl.acm.org/doi/10.1145/3544548.3580822},
   isbn = {9781450394215},
   booktitle = {Proceedings of the SIGCHI Conference on Human Factors in Computing Systems (CHI 2023)},
   month = {4},
   pages = {16},
   publisher = {Association for Computing Machinery},
   title = {Using Patient-Generated Data to Support Cardiac Rehabilitation and the Transition to Self-Care},
   year = {2023},
}

@inproceedings{Kim2024,
   author = {Taewan Kim and Seolyeong Bae and Hyun Ah Kim and Su Woo Lee and Hwajung Hong and Chanmo Yang and Young Ho Kim},
   doi = {https://doi.org/10.1145/3613904.3642937},
   isbn = {9798400703300},
   booktitle = {Proceedings of the SIGCHI Conference on Human Factors in Computing Systems (CHI 2024)},
   month = {5},
   publisher = {Association for Computing Machinery},
   title = {MindfulDiary: Harnessing Large Language Model to Support Psychiatric Patients' Journaling},
   year = {2024},
}

@inproceedings{Grimme2024,
   author = {Sophie Grimme and Susanna Marie Spoerl and Susanne Boll and Marion Koelle},
   doi = {https://doi.org/10.1145/3613904.3642851},
   issue = {24},
   booktitle = {Proceedings of the SIGCHI Conference on Human Factors in Computing Systems (CHI 2024)},
   month = {5},
   publisher = {Association for Computing Machinery},
   title = {My Data, My Choice, My Insights: Women's Requirements when Collecting, Interpreting and Sharing their Personal Health Data},
   volume = {18},
   year = {2024},
}

@article{Lai2017,
   author = {A. M. Lai and P. Y.S. Hsueh and Y. K. Choi and R. R. Austin},
   doi = {10.15265/IY-2017-016},
   issn = {23640502},
   issue = {1},
   journal = {Yearbook of Medical Informatics},
   month = {8},
   pages = {152},
   pmid = {29063559},
   publisher = {Thieme Medical Publishers},
   title = {Present and Future Trends in Consumer Health Informatics and Patient-Generated Health Data},
   volume = {26},
   url = {/pmc/articles/PMC6239232/ /pmc/articles/PMC6239232/?report=abstract https://www.ncbi.nlm.nih.gov/pmc/articles/PMC6239232/},
   year = {2017},
}

@article{Zhang2023,
   author = {Alice Qian Zhang and Ashlee Milton and Stevie Chancellor},
   doi = {10.1145/3584931.3607013},
   isbn = {9798400701290},
   journal = {Proceedings of the ACM Conference on Computer Supported Cooperative Work, CSCW Companion},
   month = {10},
   pages = {149-153},
   publisher = {Association for Computing Machinery},
   title = {\#Pragmatic or \#Clinical: Analyzing TikTok Mental Health Videos},
   url = {https://dl.acm.org/doi/10.1145/3584931.3607013},
   year = {2023},
}

@article{Wang2023,
   author = {Yihe Wang and Kathryn E. Ringland},
   doi = {10.1145/3584931.3606995},
   isbn = {9798400701290},
   journal = {Proceedings of the ACM Conference on Computer Supported Cooperative Work, CSCW Companion},
   month = {10},
   pages = {254-258},
   publisher = {Association for Computing Machinery},
   title = {Weaving Autistic Voices on TikTok: Utilizing Co-Hashtag Networks for Netnography},
   url = {https://dl.acm.org/doi/10.1145/3584931.3606995},
   year = {2023},
}

@inproceedings{Mittal2023,
   author = {Shravika Mittal and Munmun De Choudhury},
   doi = {https://doi.org/10.1145/3544548.3580834},
   isbn = {9781450394215},
   booktitle = {Proceedings of the SIGCHI Conference on Human Factors in Computing Systems (CHI 2023)},
   month = {4},
   pages = {19},
   publisher = {Association for Computing Machinery},
   title = {Moral Framing of Mental Health Discourse and Its Relationship to Stigma: A Comparison of Social Media and News},
   year = {2023},
}

@inproceedings{Young2019,
   author = {Alyson L. Young and Andrew D. Miller},
   doi = {10.1145/3290605.3300359},
   isbn = {9781450359702},
   booktitle = {Proceedings of the SIGCHI Conference on Human Factors in Computing Systems (CHI 2019)},
   month = {5},
   publisher = {Association for Computing Machinery},
   title = {"This Girl is On Fire": Sensemaking in An Online Health Community for Vulvodynia},
   url = {https://dl.acm.org/doi/10.1145/3290605.3300359},
   year = {2019},
}

@inproceedings{Huh2015,
   author = {Jina Huh},
   doi = {10.1145/2675133.2675259},
   isbn = {9781450329224},
   booktitle = {Proceedings of the ACM Conference on Computer Supported Cooperative Work (CSCW 2015)},
   month = {2},
   pages = {1488-1499},
   publisher = {Association for Computing Machinery, Inc},
   title = {Clinical Questions in Online Health Communities: The Case of "See Your Doctor" Threads},
   url = {https://dl.acm.org/doi/10.1145/2675133.2675259},
   year = {2015},
}

@inproceedings{Rubya2017,
   author = {Sabirat Rubya and Svetlana Yarosh},
   doi = {https://dl.acm.org/doi/10.1145/2998181.2998246},
   isbn = {9781450343350},
   booktitle = {Proceedings of the ACM Conference on Computer Supported Cooperative Work (CSCW 2017)},
   month = {2},
   pages = {1454-1469},
   publisher = {Association for Computing Machinery},
   title = {Video-Mediated Peer Support in an Online Community for Recovery from Substance Use Disorders},
   year = {2017},
}

@inproceedings{Chung2016,
   author = {Chia Fang Chung and Kristin Dew and Allison Cole and Jasmine Zia and James Fogarty and Julie A. Kientz and Sean A. Munson},
   doi = {10.1145/2818048.2819926},
   isbn = {9781450335928},
   booktitle = {Proceedings of the ACM Conference on Computer Supported Cooperative Work (CSCW 2016)},
   month = {2},
   pages = {770-786},
   publisher = {Association for Computing Machinery},
   title = {Boundary Negotiating Artifacts in Personal Informatics: Patient-Provider Collaboration with Patient-Generated Data},
   volume = {27},
   url = {https://dl.acm.org/doi/10.1145/2818048.2819926},
   year = {2016},
}

@article{Figueiredo2020,
   author = {Mayara Costa Figueiredo and Yunan Chen},
   doi = {10.1561/1100000080},
   isbn = {9781680835762},
   issn = {1551-3955},
   issue = {3},
   journal = {Foundations and Trends® in Human–Computer Interaction},
   month = {4},
   pages = {165-297},
   publisher = {Now Publishers, Inc.},
   title = {Patient-Generated Health Data: Dimensions, Challenges, and Open Questions},
   volume = {13},
   url = {http://dx.doi.org/10.1561/1100000080},
   year = {2020},
}

@article{Sinha2021,
   author = {Shikha Sinha and Raghuveer Puttagunta and Jennifer Vodzak},
   doi = {10.1001/JAMAPEDIATRICS.2021.2077},
   issn = {2168-6203},
   issue = {10},
   journal = {JAMA Pediatrics},
   month = {10},
   pages = {997-998},
   pmid = {34309649},
   publisher = {American Medical Association},
   title = {Interoperability and Information-Blocking Rules: Implications for Pediatric and Adolescent Health Care Professionals},
   volume = {175},
   url = {https://jamanetwork.com/journals/jamapediatrics/fullarticle/2782327},
   year = {2021},
}

@article{Ferreira2007,
   author = {Ana Ferreira and Ana Correia and Ana Silva and Ana Corte and Ana Pinto and Ana Saavedra and Ana Luís Pereira and Ana Filipa Pereira and Ricardo Cruz-Correia and Luís Filipe Antunes},
   journal = {Studies in Health Technology and Informatics},
   pages = {77-90},
   publisher = {IOS Press},
   title = {Why Facilitate Patient Access to Medical Records},
   volume = {127},
   year = {2007},
}

@inproceedings{Mosaly2016,
   author = {Prithima Reddy Mosaly and Lukasz Mazur and Lawrence B. Marks},
   doi = {10.1145/2854946.2854985},
   isbn = {9781450337519},
   booktitle = {Proceedings of the 2016 ACM Conference on Human Information Interaction and Retrieval (CHIIR 2016)},
   month = {3},
   pages = {313-316},
   publisher = {Association for Computing Machinery, Inc},
   title = {Usability Evaluation of Electronic Health Record System (EHRs) Using Subjective and Objective Measures},
   url = {https://dl.acm.org/doi/10.1145/2854946.2854985},
   year = {2016},
}

@inproceedings{Yoo2019,
   author = {Dong Whi Yoo and Munmun De Choudhury},
   doi = {10.1145/3329189.3329200},
   isbn = {9781450361262},
   issn = {21531633},
   booktitle = {Proceedings of the International Conference on Pervasive Computing Technologies for Healthcare (PervasiveHealth 2019)},
   month = {5},
   pages = {61-70},
   publisher = {ICST},
   title = {Designing Dashboard for Campus Stakeholders to Support College Student Mental Health},
   url = {https://dl.acm.org/doi/10.1145/3329189.3329200},
   year = {2019},
}

@inproceedings{Murphy2017,
   author = {Alison R. Murphy and Madhu C. Reddy},
   doi = {10.1145/2998181.2998315},
   isbn = {9781450343350},
   booktitle = {Proceedings of the ACM Conference on Computer Supported Cooperative Work (CSCW 2017)},
   month = {2},
   pages = {1646-1660},
   publisher = {Association for Computing Machinery},
   title = {Ambiguous Accountability: The Challenges of Identifying and Managing Patient-Related Information Problems in Collaborative Patient-Care Teams},
   url = {https://dl.acm.org/doi/10.1145/2998181.2998315},
   year = {2017},
}

@article{Marathe2021,
   author = {Megh Marathe and Yoonseon Yi and Chia Hsuan Su and Ting Wei Chang and Gabriela Marcu},
   doi = {10.1145/3476043},
   issn = {25730142},
   issue = {CSCW2},
   journal = {Proceedings of the ACM on Human-Computer Interaction},
   month = {10},
   pages = {24},
   publisher = {ACMPUB27New York, NY, USA},
   title = {Tedious Versus Taxing: The Nature of Work in a Behavioral Health Context},
   volume = {5},
   url = {https://dl.acm.org/doi/10.1145/3476043},
   year = {2021},
}

@article{Lv2017,
   author = {Nan Lv and Lan Xiao and Martha L. Simmons and Lisa G. Rosas and Albert Chan and Martin Entwistle},
   doi = {10.2196/jmir.7831},
   issn = {14388871},
   issue = {9},
   journal = {Journal of Medical Internet Research},
   month = {9},
   pages = {e7831},
   pmid = {28928111},
   publisher = {JMIR Publications Inc.},
   title = {Personalized Hypertension Management Using Patient-Generated Health Data Integrated With Electronic Health Records (EMPOWER-H): Six-Month Pre-Post Study},
   volume = {19},
   url = {https://www.jmir.org/2017/9/e311},
   year = {2017},
}

@inproceedings{Iott2019,
   author = {Bradley Iott and Tanner Caverly and Astrid Fishstrom and Darren King and George Meng and Allen Flynn},
   doi = {10.1145/3329189.3329198},
   isbn = {9781450361262},
   issn = {21531633},
   booktitle = {Proceedings of the International Conference on Pervasive Computing Technologies for Healthcare (PervasiveHealth 2019)},
   month = {5},
   pages = {21-30},
   publisher = {ICST},
   title = {Clinician Perspectives on the User Experience, Configuration, and Scope of Use of a Patient Reported Outcomes (PRO) Dashboard},
   url = {https://dl.acm.org/doi/10.1145/3329189.3329198},
   year = {2019},
}

@inproceedings{Pater2024,
   author = {Jessica Pater and Shaan Chopra and Juliette Zaccour and Jeanne Carroll and Fayika Farhat Nova and Tammy Toscos and Shion Guha and Fen Lei Chang},
   doi = {https://dl.acm.org/doi/10.1145/3613904.3642827},
   isbn = {9798400703300},
   booktitle = {Proceedings of the SIGCHI Conference on Human Factors in Computing Systems (CHI 2024)},
   month = {5},
   pages = {21},
   publisher = {Association for Computing Machinery},
   title = {Charting the COVID Long Haul Experience-A Longitudinal Exploration of Symptoms, Activity, and Clinical Adherence},
   url = {https://dl.acm.org/doi/10.1145/3613904.3642827},
   year = {2024},
}

@inproceedings{Veinot2010,
   author = {Tiffany C. Veinot and Kai Zheng and Julie C. Lowery and Maria Souden and Rosalind Keith},
   doi = {10.1145/1882992.1883026},
   isbn = {9781450300308},
   booktitle = {Proceedings of the 1st ACM International Health Informatics Symposium (IHI 2010)},
   pages = {240-249},
   title = {Using Electronic Health Record Systems in Diabetes Care: Emerging Practices},
   url = {https://dl.acm.org/doi/10.1145/1882992.1883026},
   year = {2010},
}

@article{Caine2013,
   author = {Kelly Caine and Rima Hanania},
   doi = {10.1136/amiajnl-2012-001023},
   issn = {1527974X},
   issue = {1},
   journal = {Journal of the American Medical Informatics Association},
   month = {1},
   pages = {7-15},
   pmid = {23184192},
   publisher = {BMJ Publishing Group},
   title = {Patients Want Granular Privacy Control over Health Information in Electronic Medical Records},
   volume = {20},
   url = {https://dx.doi.org/10.1136/amiajnl-2012-001023},
   year = {2013},
}

@article{Bourgeois2008,
   author = {Fabienne C. Bourgeois and Patrick L. Taylor and S. Jean Emans and Daniel J. Nigrin and Kenneth D. Mandl},
   doi = {10.1197/JAMIA.M2865},
   issn = {10675027},
   issue = {6},
   journal = {Journal of the American Medical Informatics Association : JAMIA},
   month = {11},
   pages = {737},
   pmid = {18755989},
   publisher = {Oxford University Press},
   title = {Whose Personal Control? Creating Private, Personally Controlled Health Records for Pediatric and Adolescent Patients},
   volume = {15},
   url = {/pmc/articles/PMC2585529/ /pmc/articles/PMC2585529/?report=abstract https://www.ncbi.nlm.nih.gov/pmc/articles/PMC2585529/},
   year = {2008},
}

@article{Weitzman2012,
   author = {Elissa R. Weitzman and Skyler Kelemen and Liljana Kaci and Kenneth D. Mandl},
   doi = {10.1186/1472-6947-12-39},
   issn = {14726947},
   journal = {BMC Medical Informatics and Decision Making},
   pages = {39},
   pmid = {22616619},
   publisher = {BMC},
   title = {Willingness to Share Personal Health Record Data for Care Improvement and Public Health: a Survey of Experienced Personal Health Record Users},
   volume = {12},
   url = {/pmc/articles/PMC3403895/ /pmc/articles/PMC3403895/?report=abstract https://www.ncbi.nlm.nih.gov/pmc/articles/PMC3403895/},
   year = {2012},
}

@article{de_Man2023,
   author = {Yvonne de Man and Yvonne Wieland-Jorna and Bart Torensma and Koos de Wit and Anneke L. Francke and Mariska G. Oosterveld-Vlug and Robert A. Verheij},
   doi = {10.2196/42131},
   issn = {14388871},
   journal = {Journal of Medical Internet Research},
   pmid = {36853745},
   publisher = {JMIR Publications Inc.},
   title = {Opt-In and Opt-Out Consent Procedures for the Reuse of Routinely Recorded Health Data in Scientific Research and Their Consequences for Consent Rate and Consent Bias: Systematic Review},
   volume = {25},
   url = {/pmc/articles/PMC10015347/ /pmc/articles/PMC10015347/?report=abstract https://www.ncbi.nlm.nih.gov/pmc/articles/PMC10015347/},
   year = {2023},
}

@article{Junghans2005,
   author = {Cornelia Junghans and Gene Feder and Harry Hemingway and Adam Timmis and Melvyn Jones},
   doi = {10.1136/BMJ.38583.625613.AE},
   issn = {0959-8138},
   issue = {7522},
   journal = {BMJ},
   month = {10},
   pages = {940},
   pmid = {16157604},
   publisher = {British Medical Journal Publishing Group},
   title = {Recruiting Patients to Medical Research: Double Blind Randomised Trial of “Opt-In” versus “Opt-Out” Strategies},
   volume = {331},
   url = {https://www.bmj.com/content/331/7522/940 https://www.bmj.com/content/331/7522/940.abstract},
   year = {2005},
}

@article{Lanier2018,
   author = {Cédric Lanier and Bernard Cerutti and Melissa Dominicé Dao and Patricia Hudelson and Noëlle Junod Perron},
   doi = {10.2147/IJGM.S178672},
   issn = {11787074},
   journal = {International Journal of General Medicine},
   pages = {393-398},
   publisher = {Dove Medical Press Ltd.},
   title = {What Factors Influence the Use of Electronic Health Records During the First 10 Minutes of the Clinical Encounter?},
   volume = {11},
   url = {https://www.tandfonline.com/action/journalInformation?journalCode=dijg20},
   year = {2018},
}

@article{Dagliati2021,
   author = {Arianna Dagliati and Alberto Malovini and Valentina Tibollo and Riccardo Bellazzi},
   doi = {10.1093/BIB/BBAA418},
   issn = {14774054},
   issue = {2},
   journal = {Briefings in Bioinformatics},
   month = {3},
   pages = {812-822},
   pmid = {33454728},
   publisher = {Oxford Academic},
   title = {Health Informatics and EHR to Support Clinical Research in the COVID-19 Pandemic: an Overview},
   volume = {22},
   url = {https://dx.doi.org/10.1093/bib/bbaa418},
   year = {2021},
}

@article{Majnarić2021,
   author = {Ljiljana Trtica Majnarić and František Babič and Shane O’sullivan and Andreas Holzinger},
   doi = {10.3390/JCM10040766},
   issn = {2077-0383},
   issue = {4},
   journal = {Journal of Clinical Medicine 2021, Vol. 10, Page 766},
   month = {2},
   pages = {766},
   publisher = {Multidisciplinary Digital Publishing Institute},
   title = {AI and Big Data in Healthcare: Towards a More Comprehensive Research Framework for Multimorbidity},
   volume = {10},
   url = {https://www.mdpi.com/2077-0383/10/4/766/htm https://www.mdpi.com/2077-0383/10/4/766},
   year = {2021},
}

@article{Ross2014,
   author = {M. K. Ross and W. Wei and L. Ohno-Machado},
   doi = {10.15265/IY-2014-0003},
   issn = {23640502},
   issue = {1},
   journal = {Yearbook of Medical Informatics},
   month = {8},
   pages = {97},
   pmid = {25123728},
   publisher = {Thieme Medical Publishers},
   title = {“Big Data” and the Electronic Health Record},
   volume = {9},
   url = {/pmc/articles/PMC4287068/ /pmc/articles/PMC4287068/?report=abstract https://www.ncbi.nlm.nih.gov/pmc/articles/PMC4287068/},
   year = {2014},
}

@article{Abul-Husn2019,
   author = {Noura S. Abul-Husn and Eimear E. Kenny},
   doi = {10.1016/J.CELL.2019.02.039},
   issn = {10974172},
   issue = {1},
   journal = {Cell},
   month = {3},
   pages = {58},
   pmid = {30901549},
   publisher = {NIH Public Access},
   title = {Personalized Medicine and the Power of Electronic Health Records},
   volume = {177},
   url = {/pmc/articles/PMC6921466/ /pmc/articles/PMC6921466/?report=abstract https://www.ncbi.nlm.nih.gov/pmc/articles/PMC6921466/},
   year = {2019},
}

@inproceedings{Tang2015,
   author = {Charlotte Tang and Yunan Chen and Bryan Semaan and Jahmeilah Roberson},
   doi = {10.1145/2675133.2675277},
   isbn = {9781450329224},
   booktitle = {Proceedings of the ACM Conference on Computer Supported Cooperative Work (CSCW 2015)},
   month = {2},
   pages = {649-661},
   publisher = {Association for Computing Machinery, Inc},
   title = {Restructuring Human Infrastructure: The Impact of HER Deployment in a Volunteer-Dependent Clinic},
   url = {https://dl.acm.org/doi/10.1145/2675133.2675277},
   year = {2015},
}

@inproceedings{Bossen2012,
   author = {Claus Bossen and Lotte Groth Jensen and Flemming Witt},
   doi = {10.1145/2145204.2145341},
   isbn = {9781450310864},
   booktitle = {Proceedings of the ACM Conference on Computer Supported Cooperative Work (CSCW 2012)},
   pages = {921-930},
   title = {Medical Secretaries' Care of Records: The Cooperative Work of a Non-Clinical Group},
   url = {https://dl.acm.org/doi/10.1145/2145204.2145341},
   year = {2012},
}

@inproceedings{Cajander2019,
   author = {Åsa Cajander and Christiane Grünloh},
   doi = {10.1145/3290605.3300865},
   isbn = {9781450359702},
   booktitle = {Proceedings of the SIGCHI Conference on Human Factors in Computing Systems (CHI 2019)},
   month = {5},
   publisher = {Association for Computing Machinery},
   title = {Electronic Health Records are More than a Work Tool Conflicting Needs of Direct and Indirect Stakeholders},
   url = {https://dl.acm.org/doi/10.1145/3290605.3300865},
   year = {2019},
}

@article{Weng2019,
   author = {Chunhua Weng and Carol Friedman and Casey A. Rommel and John F. Hurdle},
   doi = {10.1186/S12911-019-0778-Z},
   issn = {14726947},
   issue = {Suppl 3},
   journal = {BMC Medical Informatics and Decision Making},
   month = {4},
   pages = {4-7},
   pmid = {30943963},
   publisher = {BMC},
   title = {A Two-site Survey of Medical Center Personnel’s Willingness to Share Clinical Data for Research: Implications for Reproducible Health NLP Research},
   volume = {19},
   url = {/pmc/articles/PMC6448185/ /pmc/articles/PMC6448185/?report=abstract https://www.ncbi.nlm.nih.gov/pmc/articles/PMC6448185/},
   year = {2019},
}

@article{Kalkman2019,
   author = {Shona Kalkman and Johannes Van Delden and Amitava Banerjee and Benoît Tyl and Menno Mostert and Ghislaine Van Thiel},
   doi = {10.1136/MEDETHICS-2019-105651},
   issn = {1473-4257},
   issue = {1},
   journal = {Journal of medical ethics},
   month = {1},
   pages = {3-13},
   pmid = {31719155},
   publisher = {J Med Ethics},
   title = {Patients' and Public Views and Attitudes towards the Sharing of Health Data for Research: a Narrative Review of the Empirical Evidence},
   volume = {48},
   url = {https://pubmed.ncbi.nlm.nih.gov/31719155/},
   year = {2022},
}

@article{Grande2014,
   author = {David Grande and Nandita Mitra and Anand Shah and Fei Wan and David A. Asch},
   doi = {10.1001/JAMAINTERNMED.2013.9166},
   issn = {21686106},
   issue = {19},
   journal = {JAMA internal medicine},
   month = {10},
   pages = {1798},
   pmid = {23958803},
   publisher = {NIH Public Access},
   title = {Public Preferences about Secondary Uses of Electronic Health Information},
   volume = {173},
   url = {/pmc/articles/PMC4083587/ /pmc/articles/PMC4083587/?report=abstract https://www.ncbi.nlm.nih.gov/pmc/articles/PMC4083587/},
   year = {2013},
}

@article{Dinh-Le2019,
   author = {Catherine Dinh-Le and Rachel Chuang and Sara Chokshi and Devin Mann},
   doi = {10.2196/12861},
   issn = {22915222},
   issue = {9},
   journal = {JMIR mHealth and uHealth},
   month = {9},
   pages = {e12861},
   pmid = {31512582},
   publisher = {JMIR Publications Inc.},
   title = {Wearable Health Technology and Electronic Health Record Integration: Scoping Review and Future Directions},
   volume = {7},
   url = {https://mhealth.jmir.org/2019/9/e12861},
   year = {2019},
}

@article{Sandy2021,
   author = {Lisa Caputo Sandy and Thomas J. Glorioso and Kevin Weinfurt and Jeremy Sugarman and Pamela N. Peterson and Russell E. Glasgow and P. Michael Ho},
   doi = {10.1097/MD.0000000000028136},
   issn = {15365964},
   issue = {51},
   journal = {Medicine},
   month = {12},
   pmid = {34941059},
   publisher = {Wolters Kluwer Health},
   title = {Leave Me Out: Patients’ Characteristics and Reasons for Opting Out of a Pragmatic Clinical Trial Involving Medication Adherence},
   volume = {100},
   url = {/pmc/articles/PMC8702195/ /pmc/articles/PMC8702195/?report=abstract https://www.ncbi.nlm.nih.gov/pmc/articles/PMC8702195/},
   year = {2021},
}

@article{Benevento2023,
   author = {Marcello Benevento and Gabriele Mandarelli and Francesco Carravetta and Davide Ferorelli and Cristina Caterino and Simona Nicolì and Antonella Massari and Biagio Solarino},
   doi = {https://doi.org/10.3389/fpubh.2023.1213615},
   issn = {22962565},
   journal = {Frontiers in Public Health},
   pmid = {37546309},
   publisher = {Frontiers Media SA},
   title = {Measuring the Willingness to Share Personal Health Information: a Systematic Review},
   volume = {11},
   url = {/pmc/articles/PMC10397406/ /pmc/articles/PMC10397406/?report=abstract https://www.ncbi.nlm.nih.gov/pmc/articles/PMC10397406/},
   year = {2023},
}

@article{Kim2015,
   author = {Katherine K. Kim and Jill G. Joseph and Lucila Ohno-Machado},
   doi = {10.1093/JAMIA/OCV014},
   issn = {1067-5027},
   issue = {4},
   journal = {Journal of the American Medical Informatics Association},
   month = {7},
   pages = {821-830},
   pmid = {25829461},
   publisher = {Oxford Academic},
   title = {Comparison of Consumers’ Views on Electronic Data Sharing for Healthcare and Research},
   volume = {22},
   url = {https://dx.doi.org/10.1093/jamia/ocv014},
   year = {2015},
}

@article{Morse2023,
   author = {Brad Morse and Katherine K. Kim and Zixuan Xu and Cynthia G. Matsumoto and Lisa M. Schilling and Lucila Ohno-Machado and Selene S. Mak and Michelle S. Keller},
   doi = {10.1093/JAMIA/OCAD058},
   issn = {1527974X},
   issue = {6},
   journal = {Journal of the American Medical Informatics Association},
   month = {5},
   pages = {1137-1149},
   pmid = {37141581},
   publisher = {Oxford Academic},
   title = {Patient and Researcher Stakeholder Preferences for Use of Electronic Health Record Data: a Qualitative Study to Guide the Design and Development of a Platform to Honor Patient Preferences},
   volume = {30},
   url = {https://dx.doi.org/10.1093/jamia/ocad058},
   year = {2023},
}

@misc{Epic2024,
   author = {Epic},
   title = {MyChart | Powered by Epic},
   url = {https://www.mychart.org/},
   year = {2024},
}

@inproceedings{Lu2024b,
   author = {Xi Lu and Jacquelyn E. Powell and Elena Agapie and Yunan Chen and Daniel A. Epstein},
   city = {New York, NY, USA},
   doi = {https://doi.org/10.1145/3613904.3642652},
   journal = {Proceedings of the SIGCHI Conference on Human Factors in Computing Systems (CHI 2024)},
   title = {Unpacking the Lived Experience of Collaborative Pregnancy Tracking},
   year = {2024},
}

@article{Pina2020,
   author = {Laura Pina and Sang Wha Sien and Clarissa Song and Teresa M. Ward and James Fogarty and Sean A. Munson and Julie A. Kientz},
   doi = {10.1145/3392882},
   issn = {25730142},
   issue = {CSCW1},
   journal = {Proceedings of the ACM on Human-Computer Interaction},
   month = {5},
   pages = {25},
   publisher = {
		ACM
		PUB27
		New York, NY, USA
	},
   title = {DreamCatcher: Exploring How Parents and School-Age Children can Track and Review Sleep Information Together},
   volume = {4},
   year = {2020},
}

@inproceedings{Epstein2013,
   author = {Daniel A. Epstein and Alan Borning and James Fogarty},
   doi = {10.1145/2493432.2493433},
   isbn = {9781450317702},
   journal = {Proceedings of the International Conference on Ubiquitous Computing (UbiComp 2013)},
   pages = {489-498},
   title = {Fine-Grained Sharing of Sensed Physical Activity: A Value Sensitive Approach},
   year = {2013},
}

@inproceedings{Sepehri2023,
   author = {Katayoun Sepehri and Liisa Holsti and Sara Niasati and Vita Chan and Karon E. MacLean},
   doi = {10.1145/3544548.3581459},
   isbn = {9781450394215},
   issue = {23},
   journal = {Proceedings of the SIGCHI Conference on Human Factors in Computing Systems (CHI 2023)},
   month = {4},
   publisher = {Association for Computing Machinery},
   title = {Beyond the Bulging Binder: Family-Centered Design of a Digital Health Information Management System for Caregivers of Children Living with Health Complexity},
   year = {2023},
}

@inproceedings{Gui2017a,
   author = {Xinning Gui and Yu Chen and Clara Caldeira and Dan Xiao and Yunan Chen},
   doi = {10.1145/3025453.3025654},
   isbn = {9781450346559},
   journal = {Proceedings of the SIGCHI Conference on Human Factors in Computing Systems (CHI 2017)},
   month = {5},
   pages = {1647-1659},
   title = {When Fitness Meets Social Networks: Investigating Fitness Tracking and Social Practices on WeRun},
   volume = {2017-May},
   url = {https://doi.org/10.1145/3025453.3025654},
   year = {2017},
}

@inproceedings{Hong2018,
   author = {Matthew K. Hong and Udaya Lakshmi and Thomas A. Olson and Lauren Wilcox},
   city = {New York, New York, USA},
   doi = {10.1145/3173574.3174050},
   isbn = {9781450356206},
   journal = {Conference on Human Factors in Computing Systems - Proceedings},
   month = {4},
   pages = {1-13},
   publisher = {Association for Computing Machinery},
   title = {Visual ODLs: Co-Designing Patient-Generated Observations of Daily Living to Support Data-Driven Conversations in Pediatric Care},
   volume = {2018-April},
   url = {http://dl.acm.org/citation.cfm?doid=3173574.3174050},
   year = {2018},
}

@inproceedings{Mehrnezhad2022,
   author = {Maryam Mehrnezhad and Laura Shipp and Teresa Almeida and Ehsan Toreini},
   doi = {10.1145/3549015.3554204},
   isbn = {9781450397001},
   journal = {Proceedings of the European Symposium on Usable Security (EuroUSEC 2022)},
   month = {9},
   pages = {145-150},
   publisher = {Association for Computing Machinery},
   title = {Vision: Too Little too Late? Do the Risks of FemTech already Outweigh the Benefits?},
   year = {2022},
}

@inproceedings{Yamashita2017,
   author = {Naomi Yamashita and Hideaki Kuzuoka and Keiji Hirata and Takashi Kudo and Eiji Aramaki and Kazuki Hattori},
   doi = {10.1145/3025453.3025843},
   isbn = {9781450346559},
   journal = {Proceedings of the SIGCHI Conference on Human Factors in Computing Systems (CHI 2017)},
   month = {5},
   pages = {158-169},
   publisher = {Association for Computing Machinery},
   title = {Changing Moods: How Manual Tracking by Family Caregivers Improves Caring and Family Communication},
   volume = {2017-May},
   year = {2017},
}

@inproceedings{Massimi2014,
   author = {Michael Massimi and Jackie Bender and Holly O. Witteman and Osman Hassan Ahmed},
   doi = {10.1145/2531602.2531622},
   isbn = {9781450325400},
   journal = {Proceedings of the ACM Conference on Computer Supported Cooperative Work (CSCW 2014)},
   pages = {1491-1501},
   publisher = {Association for Computing Machinery},
   title = {Life Transitions and Online Health Communities: Reflecting on Adoption, Use, and Disengagement},
   year = {2014},
}

@article{Jo2022,
   author = {Eunkyung Jo and Seora Park and Hyeonseok Bang and Youngeun Hong and Yeni Kim and Jungwon Choi and Bung Nyun Kim and Daniel A. Epstein and Hwajung Hong},
   doi = {10.1145/3512939},
   isbn = {25730142/2022/4},
   issn = {25730142},
   issue = {CSCW1},
   journal = {Proceedings of the ACM on Human-Computer Interaction},
   month = {4},
   pages = {92},
   publisher = {
		ACM
		PUB27
		New York, NY, USA
	},
   title = {GeniAuti: Toward Data-Driven Interventions to Challenging Behaviors of Autistic Children through Caregivers' Tracking},
   volume = {6},
   year = {2022},
}

@inproceedings{Pina2017,
   author = {Laura R Pina and Sang-Wha Sien and Teresa Ward and Jason C Yip and Sean A Munson and James Fogarty and Julie A Kientz},
   city = {New York, NY, USA},
   doi = {10.1145/2998181.2998362},
   isbn = {9781450343350},
   journal = {Proceedings of the ACM Conference on Computer Supported Cooperative Work (CSCW 2017)},
   pages = {2300-2315},
   publisher = {ACM},
   title = {From Personal Informatics to Family Informatics: Understanding Family Practices around Health Monitoring},
   year = {2017},
}

@inproceedings{Diethei2021,
author = {Diethei, Daniel and Niess, Jasmin and Stellmacher, Carolin and Stefanidi, Evropi and Sch{\"{o}}ning, Johannes},
booktitle = {Proceedings of the SIGCHI Conference on Human Factors in Computing Systems (CHI 2021)},
doi = {10.1145/3411764.3445665},
month = {May},
title = {{Sharing Heartbeats: Motivations of Citizen Scientists in Times of Crises}},
year = {2021}
}

@inproceedings{Arzt2021,
   author = {Steven Arzt and Andreas Poller and Gisela Vallejo},
   doi = {10.1145/3411763.3451631},
   journal = {Proceedings of the SIGCHI Conference on Human Factors in Computing Systems (CHI 2021)},
   month = {5},
   title = {Tracing Contacts with Mobile Phones to Curb the Pandemic: Topics and Stances in People's Online Comments about the Official German Contact-Tracing App},
   year = {2021},
}

@article{Lim2019,
   doi = {10.1145/3328928},
   issn = {2474-9567},
   issue = {2},
   journal = {Proceedings of the ACM on Interactive, Mobile, Wearable and Ubiquitous Technologies},
   month = {6},
   pages = {1-46},
   publisher = {Association for Computing Machinery (ACM)},
   title = {How Does a Nation Walk? Interpreting Large-Scale Step Count Activity with Weekly Streak Patterns},
   volume = {3},
   url = {https://dl.acm.org/doi/10.1145/3328928},
   year = {2019},
}

@incollection{Braun2012,
author = {Braun, Virginia and Clarke, Victoria},
booktitle = {APA handbook of research methods in psychology, Vol 2: Research designs: Quantitative, qualitative, neuropsychological, and biological.},
doi = {10.1037/13620-004},
month = {mar},
pages = {57--71},
publisher = {American Psychological Association},
title = {{Thematic Analysis.}},
year = {2012}
}

@inproceedings{Chung2017,
   author = {Chia-Fang Chung and Elena Agapie and Jessica Schroeder and Sonali Mishra and James Fogarty and Sean A Munson},
   city = {New York, NY, USA},
   doi = {10.1145/3025453.3025747},
   isbn = {9781450346559},
   journal = {Proceedings of the SIGCHI Conference on Human Factors in Computing Systems (CHI 2017)},
   publisher = {ACM},
   title = {When Personal Tracking Becomes Social: Examining the Use of Instagram for Healthy Eating},
   url = {http://doi.org/10.1145/3025453.3025747},
   year = {2017},
}

@inproceedings{Epstein2020,
   author = {Daniel A. Epstein and Clara Caldeira and Mayara Costa Figueiredo and Xi Lu and Lucas M. Silva and Lucretia Williams and Jong Ho Lee and Qingyang Li and Simran Ahuja and Qiuer Chen and Payam Dowlatyari and Craig Hilby and Sazeda Sultana and Elizabeth V. Eikey and Yunan Chen},
   doi = {10.1145/3432231},
   issn = {24749567},
   issue = {4},
   journal = {Proceedings of the ACM on Interactive, Mobile, Wearable and Ubiquitous Technologies (IMWUT 2020)},
   month = {12},
   pages = {126},
   publisher = {Association for Computing Machinery},
   title = {Mapping and Taking Stock of the Personal Informatics Literature},
   volume = {4},
   url = {https://doi.org/10.1145/3432231},
   year = {2020},
}

@inproceedings{Lu2021a,
   author = {Xi Lu and Hwajung Hong and Tera L. Reynolds and Xinru Page and Daniel A. Epstein and Eunkyung Jo and Yunan Chen},
   city = {New York, NY, USA},
   doi = {10.1145/3411764.3445669},
   journal = {Proceedings of the SIGCHI Conference on Human Factors in Computing Systems (CHI 2021)},
   title = {Comparing Perspectives Around Human and Technology Support for Contact Tracing},
   year = {2021},
}

@inproceedings{Figueiredo2021,
   author = {Mayara Costa Figueiredo and Yunan Chen},
   doi = {10.1145/3411764.3445189},
   isbn = {9781450380966},
   journal = {Proceedings of the SIGCHI Conference on Human Factors in Computing Systems (CHI 2021)},
   month = {5},
   pages = {17},
   publisher = {Association for Computing Machinery},
   title = {Health Data in Fertility Care: An Ecological Perspective},
   year = {2021},
}

@inproceedings{Gui2017,
   author = {Xinning Gui and Yu Chen and Yubo Kou and Kathleen H. Pine and Yunan Chen},
   doi = {10.1145/3134685},
   isbn = {10.1145/3134685},
   issn = {25730142},
   issue = {CSCW},
   journal = {Proceedings of the ACM Conference on Computer Supported Cooperative Work (CSCW 2017)},
   month = {11},
   pages = {19},
   publisher = {Association for Computing Machinery},
   title = {Investigating Support Seeking from Peers for Pregnancy in Online Health Communities},
   volume = {1},
   year = {2017},
}

@inproceedings{Murnane2018,
   author = {Elizabeth L. Murnane and Tara G. Walker and Beck Tench and Stephen Voida and Jaime Snyder},
   doi = {10.1145/3274396},
   issn = {25730142},
   issue = {CSCW},
   journal = {Proceedings of the ACM Conference on Computer Supported Cooperative Work (CSCW 2018)},
   month = {11},
   title = {Personal Informatics in Interpersonal Contexts: Towards the Design of Technology that Supports the Social Ecologies of Long-Term Mental Health Management},
   volume = {2},
   year = {2018},
}

@inproceedings{Lu2022,
   author = {Xi Lu and Eunkyung Jo and Seora Park and Hwajung Hong and Yunan Chen and Daniel A. Epstein},
   doi = {https://doi.org/10.1145/3555569},
   journal = {Proceedings of the ACM Conference on Computer Supported Cooperative Work (CSCW 2022)},
   title = {Understanding Cultural Influence on Perspectives Around Contact Tracing Strategies},
   year = {2022},
}


\end{document}